\documentclass[%
 reprint,
nofootinbib,
 amsmath,amssymb,
 aps,
floatfix,
longbibliography,
]{revtex4-2}

\usepackage[english]{babel}
\usepackage[utf8]{inputenc}
\usepackage{graphicx}
\usepackage{graphics}
\usepackage{dcolumn}
\usepackage{bm}
\usepackage[mathlines]{lineno}
\usepackage{xcolor}
\usepackage{dcolumn}

\usepackage{soul}
\newcommand{\Hl}[2][\empty]{%
    \ifx#1\empty
    \else
    \sethlcolor{#1}%
    \fi
    \hl{#2}}
\soulregister\Hl{7}
\soulregister\cite7
\soulregister\ref7
\soulregister\pageref7
\usepackage{amssymb}               
\usepackage{url}                   
\usepackage{rotating}              
\usepackage{multirow}              
\usepackage{tikz}
\DeclareUnicodeCharacter{2212}{-}

\newcommand{\degree}{\ensuremath{^\circ}}
\usepackage{pict2e} 
\usepackage[permil]{overpic}
\usepackage{multirow}
\usepackage{threeparttable}
\usepackage{supertabular}
\usepackage{footnote}
\usepackage{tablefootnote}
\usepackage{longtable} 
\usepackage{adjustbox}
\usepackage{placeins}

\begin{document}
\preprint{LHCb-implications-2}
\title{Energy deposition studies for the Upgrade II of LHCb at the CERN Large Hadron Collider}

\author{Alessia Ciccotelli}
 \email{alessia.ciccotelli@cern.ch}
\affiliation{The University of Manchester and the Cockcroft Institute, Department of Physics and Astronomy, School of Natural Sciences, Oxford Road, Manchester M13 9PL, United Kingdom
}%
\affiliation{CERN, CH 1211 Geneva 23, Switzerland}

\author{Robert B. Appleby}
\affiliation{The University of Manchester and the Cockcroft Institute, Department of Physics and Astronomy, School of Natural Sciences, Oxford Road, Manchester M13 9PL, United Kingdom
}%

\author{Francesco Cerutti}%
\affiliation{%
 CERN, CH 1211 Geneva 23, Switzerland
}%
\author{Kevin Buffet}%
\affiliation{%
 CERN, CH 1211 Geneva 23, Switzerland
}%
\author{Francois Butin}%
\affiliation{%
 CERN, CH 1211 Geneva 23, Switzerland
}%
\author{Gloria Corti}
\affiliation{%
 CERN, CH 1211 Geneva 23, Switzerland
}
\author{Luigi Salvatore Esposito}%
\affiliation{%
 CERN, CH 1211 Geneva 23, Switzerland
}%
\author{Ruben Garcia Alia}%
\affiliation{%
 CERN, CH 1211 Geneva 23, Switzerland
}%
\author{Matthias Karacson}
\affiliation{%
 CERN, CH 1211 Geneva 23, Switzerland
}
\author{Giuseppe Lerner}%
\affiliation{%
 CERN, CH 1211 Geneva 23, Switzerland
}%
\author{Daniel Prelipcean}%
\email{Technical University of Munich (TUM), Arcisstraße 21, 80333 München, Germany
}
\affiliation{%
 CERN, CH 1211 Geneva 23, Switzerland
}%
\author{Maud Wehrle}%
\affiliation{%
 CERN, CH 1211 Geneva 23, Switzerland
}%
\date{\today}

\begin{abstract}
The Upgrade II of the LHCb (Large Hadron Collider beauty) experiment is proposed to be installed during the CERN Long Shutdown 4, aiming to operate LHCb at $1.5\cdot\,10^{34}\,\mathrm{cm^{−2}\,s^{−1}}$ that is 75 times its design luminosity~\cite{Alves:1129809} and reaching an integrated luminosity of about $400\,\mathrm{fb^{−1}}$ by the end of the High Luminosity LHC era.
This increase of the data sample at LHCb is an unprecedented opportunity for heavy flavour physics measurements. A first upgrade of LHCb (Upgrade I), completed in 2022, has already implemented important changes of the LHCb detector and, for the Upgrade II, further detector improvements of the tracking system, the particle identification system and the online and trigger infrastructure are being considered.
Such a luminosity increase will have an impact not only on the LHCb detector but also on the LHC magnets, cryogenics and electronic equipment placed in the insertion region 8. In fact, the LHCb experiment was conceived to work at a much lower luminosity than ATLAS and CMS, implying minor requirements for protection of the LHC elements from the collision debris and therefore a different layout around the interaction point. The Upgrade I has already implied the installation of an absorber for the neutral particle debris (TANB). However, the luminosity target proposed for this second upgrade requires to review the layout of the entire insertion region in order to ensure safe operation of the LHC magnets and to mitigate the risk of failure of the electronic devices. The objective of this paper is to provide an overview of the implications of the Upgrade~II of LHCb in the experimental cavern and in the tunnel with a focus on the LHCb detector, electronic devices and accelerator magnets. This proves that the Upgrade II luminosity goal can be sustained with the implementation of protection systems for magnets and electronics. The electronics placed in the experimental areas and in the cavern needs to be protected to mitigate the single event effect (SEE) risk, which may imply recurring downtime of the LHC. 
On the other hand, protection for the first quadrupole of the final focus triplet and for the separation dipole are needed to prevent their quench and to reach the desired lifetime. Moreover, the normal-conducting compensators require the installation of shielding to limit their head coil degradation.  
\end{abstract} 
\maketitle

\section{\label{sec:level1}Introduction}
The High Luminosity era of the Large Hadron Collider (LHC) is expected to start operation in 2029 and reach $4000\,\mathrm{fb^{-1}}$ according to its ultimate luminosity goal~\cite{CERN_LHC_schedule}. 
The High Luminosity LHC (HL-LHC) project has been dealing with this technological and scientific challenge to extend physics discovery potential at CERN. The HL-LHC is designed to allow proton--proton collisions at a center-of-mass energy $\sqrt{s}=14\,\mathrm{TeV}$ and to deliver an instantaneous luminosity of $5-7.5\cdot 10^{34}\,cm^{-2}\,s^{-1}$ to the ATLAS and CMS experiments. This unprecedented luminosity will allow more accurate measurements of new particles and the observation of rare processes that happen below the current LHC sensitivity level. In parallel to the HL-LHC project, the upgrade of the detectors shall make the experimental apparatus capable of sustaining higher radiation levels and of analysing a larger amount of data from particles collisions.

The other LHC large experiments, LHCb and Alice, have more specialized detectors that are designed to work with lower luminosities.
This study is dedicated to the LHCb experiment, placed in the insertion region 8 (IR8) and operated at $4\cdot10^{32}\,\mathrm{cm}^{−2}\,\mathrm{s}^{−1}$ during Run~2. To push the physics programme to broader horizons, a first upgrade of the LHCb experiment (Upgrade~I) was realised during the CERN machine long shutdown from mid-2018 to mid-2022 (LS2) allowing it to operate with a significantly increased instantaneous luminosity, up to $2\cdot\,10^{33}\,\mathrm{cm^{−2}\,s^{−1}}$. Despite the considerable technological effort and promising physics results expected from the Upgrade~I of LHCb, the upgraded detector does not reach ultimate precision in many key physics observables~\cite{LHCbCollaboration:2776420} and more data would be needed to fully exploit the flavour-physics potential at the HL-LHC. Therefore, a second upgrade of the LHCb experiment (Upgrade~II) has been proposed to be implemented during the CERN machine long shutdown in 2033--2035 (LS4), aiming to increase the instantaneous luminosity by 7.5 times to a value of $1.5\cdot\,10^{34}\,\mathrm{cm^{−2}\,s^{−1}}$ and collect an integrated luminosity close to $400\,\mathrm{fb^{-1}}$ by the end of HL-LHC era. 
Concerning the impact of the proposed increase of the LHCb luminosity, the risk of magnet quench, the higher cryogenic load, the magnet lifetime, and the radiation to electronics constitute demanding challenges for the full implementation of the Upgrade~II. 
A preliminary analysis of the implications of working at high luminosity in IR8 was already released in 2018~\cite{Efthymiopoulos:2319258} to indicate the critical issues arisen from first studies.
Moreover, a study on the implications of the Upgrade~I of LHCb on the accelerator was published~\cite{PhysRevAccelBeams.26.061002} in 2023 laying the groundwork for the present paper.
The objective of this study is to provide an overview of the implications of the Upgrade~II of LHCb on the experimental cavern and the tunnel equipment. 

This paper explores all the identified issues and the state-of-the-art mitigation measures.
Section~\ref{sec:level2} describes the provisional schedule of the LHCb operation, justifying the assumptions and normalization factors used in the study. Section~\ref{sec:level3} expounds the relevant aspects of the Monte Carlo calculation, in terms of radiation source and evaluated quantities, and illustrates the geometry model of the LHC accelerator and of the LHCb detector.
An overview of the implications of Upgrade II on the LHCb detector is presented in Section~\ref{sec:level5}. Then, Section~\ref{sec:level6} shows the radiation levels in the experimental cavern and service tunnels as well as a solution to mitigate the risk of electronics failures.
Moreover, the discussion of the main issues related to the cryoload and the lifetime and quench risk of the LHC magnets is presented in Section~\ref{sec:level7}. Finally we draw our conclusions. 

We find that the proposed luminosity is achievable from a machine protection and radiation level perspective, for both the LHC machine and the upgraded detector, provided that the indicated measures (or equivalent ones) are implemented.


\section{Provisional schedule for post--Upgrade II of LHC\lowercase{b}\label{sec:level2}}

\begin{figure}[!tbh]
    \centering   
    \includegraphics*[trim=0.1cm 0.1cm 0.1cm 0.1cm, clip=true,width=0.95\columnwidth]{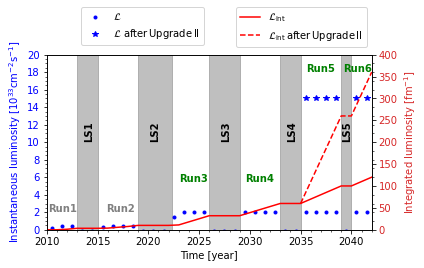}

    \caption{Luminosity performances in IR8 with and without the Upgrade II of LHCb~\cite{LHCbCollaboration:2776420,CERN_LHC_schedule,triplet}. }
    \label{fig:lumi}
\end{figure}

In mid-2022, the LHC restarted beam operation with the Run 3 phase after LS2.
During the latter, the LHC and its injector chain were consolidated and the LHCb detector underwent its first major upgrade I.
The first year of Run~3 (2022) delivered to LHCb an integrated luminosity of approximately $1\,\mathrm{fb^{-1}}$, with horizontal beam crossing. In 2023, operation started with the adoption of the external crossing on the vertical plane, as explained in detail in ~\cite{PhysRevAccelBeams.26.061002}. 
This is desirable for the LHCb physics programme because it guarantees symmetry between the two LHCb dipole magnet polarities, reducing some sources of systematic uncertainties~\cite{Albrecht:2653011}. 
As shown in Fig.~\ref{fig:lumi}, the Upgrade II of LHCb is designed to enable an increase in instantaneous luminosity from $2\cdot\,10^{33}\,\mathrm{cm^{−2}\,s^{−1}}$ to $1.5\cdot\,10^{34}\,\mathrm{cm^{−2}\,s^{−1}}$ during Run~5 and Run~6. Furthermore, with regard to the integrated luminosity LHCb would collect until the end of Run 6 $370\,\mathrm{fb^{−1}}$ in the Upgrade II scenario compared to $130\,\mathrm{fb^{−1}}$ in the baseline HL-LHC scenario. Instantaneous and cumulative radiation levels during operation scale proportionally to the instantaneous and integrated luminosity, respectively. 
 


\section{Simulation modeling of IR8 and LHC\lowercase{b} detector\label{sec:level3} }

The energy deposition in high energy collider components is determined by particle shower propagation through the materials. Monte Carlo simulations are a valuable tool to predict the implications of controlled and uncontrolled beam loss impacts on the accelerator elements, surrounding electronics and environment. In this study, radiation levels are calculated by means of the FLUKA code~\cite{FLUKA,FLUKA2021,FLUKA:2015}, which provides an accurate modeling of particle interactions over a wide energy range.

The stand-alone FLUKA model of IR8 contains a full three-dimensional description of the beam line and tunnel, as well as the LHCb cavern and detector.
The energy deposition calculations are based on realistic geometry models of magnets, collimators, absorbers, and vacuum chambers.
This detailed model of IR8 was implemented by means of auxiliary tools: a dedicated Python-based code for FLUKA geometry assembling, called Linebuilder~\cite{Mereghetti:2012zz}, and Flair, which is the User Graphical Interface of FLUKA~\cite{Vlachoudis:2749540}.

The radiation shower in IR8 is dominated by inelastic proton--proton collisions at the Interaction Point 8 (IP8), whose cross section, including diffractive events, is assumed to be $\sigma_{pp}=80\,\mathrm{mb}$ at 14 TeV center-of-mass energy \cite{ATLAS:2016ygv}.
Simulation parameters used for this study are reported in Table~\ref{sim_settings}.

\begin{table}[!hbt]
   \centering
   \caption{Simulation parameters used for the FLUKA simulation studies of the Upgrade II of LHCb}
   \begin{ruledtabular}
   \begin{tabular}{lc}
        \multicolumn{2}{c}{ \textbf{Simulation parameters}} \\ \multicolumn{2}{c}{ \textbf{HL-LHC LHCb Upgrade II}} \\
    \colrule
        \textbf{p-p collisions} & beam energy of $7\,\mathrm{TeV}$ \\ 
        \textbf{External crossing angle} & vertical plane \\ 
        \multirow{2}{10em}{\textbf{Integrated luminosity}} & Final target ${400\,\mathrm{fb^{-1}}}$  \\ 
          & Annual ${50\,\mathrm{fb^{-1}}}$ \\ 
        \textbf{Instantaneous luminosity} & $1.5\cdot 10^{34}\,\mathrm{cm^{-2}}\,\mathrm{s^{-1}}$ \\
   \end{tabular}
   \label{sim_settings}
   \end{ruledtabular}
\end{table}

With $7\,\mathrm{TeV}$ beams and a $1.2\cdot\,10^{9}\,\mathrm{s^{−1}}$ inelastic collision rate, the collision debris power towards either side of IP8 is $1.3\,\mathrm{kW}$, compared to $174\,\mathrm{W}$ that is the one during Run~3 operation with $6.8\,\mathrm{TeV}$ beams and $2\cdot 10^{33}\,\mathrm{cm^{-2}}\,\mathrm{s^{-1}}$ instantaneous luminosity.

In our study, we quantify the total power on each accelerator element, that is crucial to define the heat load to be extracted by the cryogenic system.
Moreover, we assess the energy deposition density distribution in the superconducting coils of the LHC magnets.
In fact, for the beam collider to function, it is essential that they do not switch to the normal-conducting regime. This process is called \textit{quench} and occurs when the deposited power density exceeds a certain limit, called \textit{quench limit}, that differs from magnet to magnet and is provided by calculations and/or experimental evidence. In this study, the local relevant quantity is evaluated as the peak power density averaged over the cable radial thickness of the inner coil layer with an azimuthal resolution of $2\degree$ and a longitudinal resolution of typically $10\,\mathrm{cm}$, as already done in similar studies~\cite{PhysRevAccelBeams.26.061002,Brugger:2131739}. Finally, dedicated total ionizing dose (TID) scoring is used to determine if a magnet, superconducting or normal-conducting (as the compensator dipoles~\cite{PhysRevAccelBeams.26.061002}), can sustain the desired lifetime, i.e. the integrated luminosity up to the Run~6 end. The problem is related to the coil insulator, which is particularly exposed to progressive deterioration due to radiation. In this case, the peak dose is evaluated with a finer radial resolution by means of bins not exceeding $3\,\mathrm{mm}$, which has been found to be a good compromise between accuracy maximization and computational time minimization.
Concerning the radiation levels in the tunnel and the experimental cavern areas, the quantities considered in the paper are TID, high-energy hadron equivalent (HEH) fluence $\Phi_{HEH}$ and thermal neutron equivalent (n-th) fluence $\Phi_{n-th}$. TID correlates with cumulative radiation effects limiting the electronics lifetime. In this regard, also calculations of Si-1-MeV neutron equivalent (1MeVn-eq) fluence were carried out to quantify Displacement Damage (DD) in Silicon, but the results are not included hereafter because they do not actually change the conclusions. To assess the Single Event Effects (SEE) probability, both HEH fluence and n-th fluence are considered. The HEH fluence is defined as the fluence of all hadrons with energy greater than $20\,\mathrm{MeV}$ plus the contribution of intermediate energy neutrons from $0.2\,\mathrm{MeV}$ to $20\,\mathrm{MeV}$. The n-th fluence includes neutrons at thermal energy and weighed contribution from the higher energy range. Radiation safe areas imply annual values of HEH and n-th fluence below $3\cdot10^{6}\,\mathrm{cm^{-2}}\,\mathrm{yr^{-1}}$ and $3\cdot10^{7}\,\mathrm{cm^{-2}}\,\mathrm{yr^{-1}}$, respectively~\cite{Lerner:2302154}. The FLUKA scoring of these quantities was performed on a Cartesian mesh with three-dimensional bins of $20\,\mathrm{cm}\times20\,\mathrm{cm}\times20\,\mathrm{cm}$ volume. 

\subsection{LHCb detector}
The LHCb detector is a single arm detector designed for studying particles produced by the collision in the forward direction.
Therefore, it has little to no coverage at large polar angles, but instead it is a spectrometer consisting of a dipole magnet and a series of planar sub-detectors positioned along the beam line mainly on the right side with respect to the IP (i.e., in the clockwise direction) near the beam line. Starting with the Vertex Locator (VELO) that is designed to approach and surround the IP during operation, particles created in the p--p collisions are traced through a first Ring-Imaging Cherenkov detector (RICH1), followed by the Upstream Tracker (UT). The UT is positioned just at the starting edge of the LHCb magnet. After particles traverse the dipole field, they are subsequently traced through the Scintillating Fiber detector (SciFi). The VELO, UT and SciFi were all installed during Upgrade I in LS2 to replace the tracking detectors that were in these locations in Run~1 and Run~2.
Particles are subsequently identified by the second RICH detector (RICH2), the Electromagnetic (ECAL) and Hadronic (HCAL) Calorimeters that are following, and finally by four Muon stations. These last stations are interspersed within iron shielding walls. Large shielding blocks are also present behind the last of the Muon stations in between the edge of the experimental cavern and the LHC tunnel. In its current configuration, a neutron shield made of polyethylene was placed in LS2 immediately upstream of the ECAL in order to protect the SciFi photomultipliers from neutron backsplash \cite{CalvoGomez:2268537,CalvoGomez:2869609}.

Pile-up reduction from what is offered by the LHC originates from the need to distinguish the vertices of particle decays and p--p collisions with precision in many of the LHCb physics measurements coupled with the proximity of the detector to the beam line. This proximity also poses the requirement of radiation level compatibility with reliable electronics operation.
For these reasons, 
LHCb was designed to collect lower luminosity than ATLAS and CMS, and major upgrades in key detector elements are therefore necessary for a substantial luminosity increase.

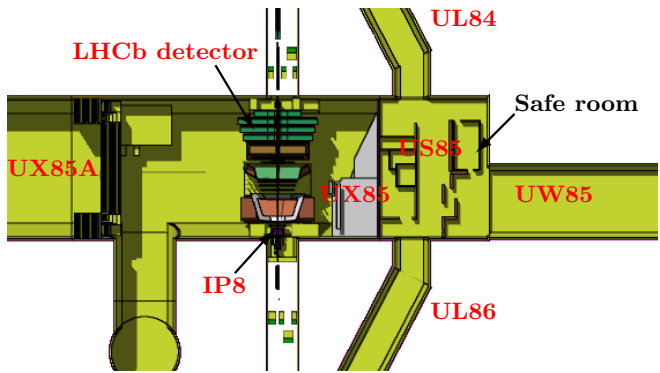
\begin{figure}[h!]
      \centering
        \begin{overpic}[trim=2.0cm 1.2cm 0.5cm 1.0cm, clip=true,width=1.0\columnwidth]{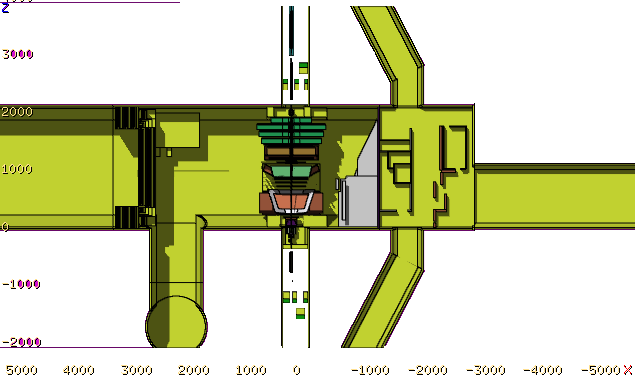}
        \put (100,480) {\textbf{\color{red} LHCb detector}}
        \put (480,260) {\textbf{\color{red} UX85}}
        \put (780,260) {\textbf{\color{red} UW85}}
        \put (0,300) {\textbf{\color{red} UX85A}}
        \put (600,330) {\textbf{\color{red} US85}}
        \put (650,530) {\textbf{\color{red} UL84}}
        \put (650,80) {\textbf{\color{red} UL86}}
        \put (300,120) {\textbf{\color{red} IP8}}
        \put(350,150){\linethickness{0.3mm}\color{black}\vector(0.9,1){60}}
        \put (780,400) {\textbf{\color{black} Safe~room}}
        \put(780,395){\linethickness{0.3mm}\color{black}\vector(-1,-1){60}}
        \put(320,470){\linethickness{0.3mm}\color{black}\vector(0.6,-1){60}}
        \end{overpic}
      \caption{Top view at beam height of the FLUKA geometry of the LHCb cavern and surrounding areas.}
      \label{fig:geo_cavern}
    \end{figure}

 Figure~\ref{fig:geo_cavern} shows the FLUKA geometry layout of the LHCb detector, the experimental cavern, LHC tunnel and surrounding areas. Cryogenic boxes are located in UX85 and electronics racks are present in US85, UL84, UL86. 

The presence of the LHCb spectrometer produces a deflection of the circulating beam on the horizontal plane. The magnet is designed to operate with two opposite polarities to collect equal amounts of integrated luminosity with either spectrometer configuration.
The LHCb spectrometer deflection is compensated by three normal-conducting dipoles: the MBXWH, placed on the left side of IP8 with a magnetic field of equal magnitude, but opposite direction, with respect to the one of the LHCb dipole, and two short dipoles (MBXWS), with opposite fields, placed on either side of IP8 close to the final focus triplet. 

IP8 is displaced by about 11 m towards IP7 to leave space for the single arm detector in the cavern, whose center is also the LHC octant center. The shift of IP8 implies a consequent displacement of the entire string of the magnets up to the recombination dipole (D2). This shift is recuperated before the dispersion suppressor (DS) causing the asymmetry of the matching section layout with respect to IP8. This, together with the presence of the injection line for the counterclockwise beam 2 on the right side of IP8, makes IR8 specific from different points of view.

%
 \begin{figure}[!tbh]
    \centering   
\includegraphics*[trim=0.0cm 1.0cm 0.0cm 0.0cm, clip=true,width=0.95\columnwidth]{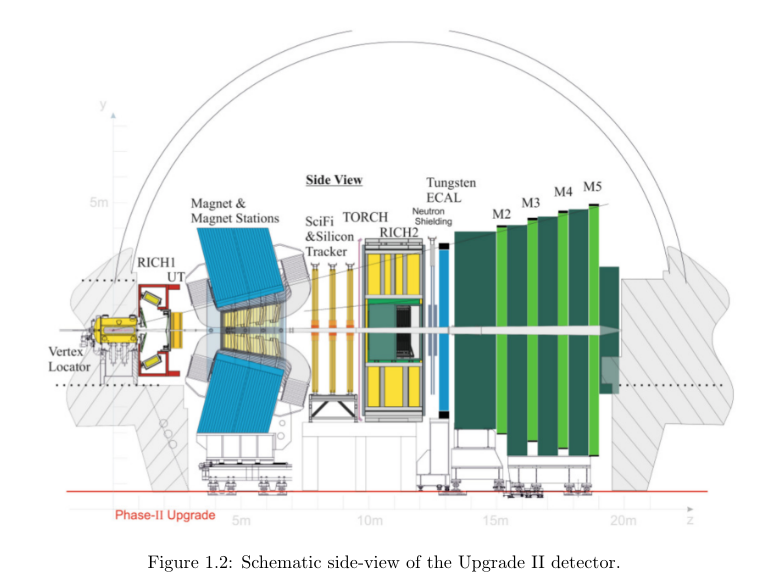}

    \caption{Schematic side-view of the Upgrade II detector extracted from the Technical Design Report (TDR) published by the LHCb Collaboration \cite{LHCbCollaboration:2776420}.}
    \label{fig:detector}
\end{figure}

\section{Implications on the LHC\lowercase{b} detector\label{sec:level5}}

Due to its unique geometry and high-precision capabilities, LHCb is positioned to potentially discover new physics that are out of reach of other experiments. LHCb already provides benchmark measurements of CP-violation parameters related to the origin of matter-antimatter asymmetry with unprecedented precision. Prospectively, several of its physics objectives are about to cross the important threshold of Standard Model predictions at the luminosity level expected after Upgrade II, promising exciting insights into beauty and charm physics, rare decays, electroweak physics, lepton flavor universality, pentaquark states and more. Furthermore, Upgrade II will also expand the scope of its physics program during heavy ion collisions as well as fixed target runs.

Optimal technological choices for Upgrade II of the LHCb detector are under active investigation within the LHCb collaboration. While the basic layout will remain the same of Run~3, various technological choices are under consideration in order to maintain the same detector performance, and in some aspects improve it, at a 7.5 times higher pile-up than that of Run~3. An average of around 40 interactions per bunch crossing will lead to much higher particle multiplicities within the acceptance of the detector. Precise reconstructions of these events to identify signatures of interesting physics observables will be critical. Most of the existing components will have to be replaced to increase their granularity and timing resolution to be able to cope with the higher particle density. A baseline configuration is described in the Upgrade II TDR \cite{LHCbCollaboration:2776420} and the corresponding schematic is shown in Figure~\ref{fig:detector}. New developments of electronics for all detectors, with added timing information and higher radiation tolerance are needed to continue to perform  particle identification and high-precision measurements at the current level in the HL-LHC environment. Innovative technological solutions for detectors will be required to meet these expectations. For example, the Vertex Locator (VELO) will transform into a true 4D tracking detector as it must be capable of independent reconstruction of primary vertices with a timing precision better than 50 ps. In addition, new types of detectors are in development to further exploit yet unexplored phase space and increase the physics reach of the experiment. New scintillating fiber stations covering the sidewalls of the LHCb dipole will extend the low-momentum tracking capabilities of LHCb. An additional silicon pixel (Mighty) Tracker close to the beam pipe will enhance the sensitivity in the area currently covered by the Scintillating Fiber Tracker (SciFi). A new high precision time of flight detector (TORCH) is envisioned to complement the design and expand hadron identification by measuring internally reflected Cherenkov radiation produced in a thin quartz plane within a time resolution of a few tens of ps. The central area of the Electromagnetic Calorimeter (ECAL) will be modified to increase granularity and enhance the timing resolution to a few tens of ps per particle, which is crucial to suppress combinatorial background at the expected pile-up. The increase in luminosity will also necessitate new shielding solutions for the MUON system in order to keep the most inner region fully operational after Upgrade II. In this context, the complete replacement of the current Hadron Calorimeter (HCAL) by a shielding solution is under consideration. The technological solutions adopted by the various sub-detectors will be detailed upon in dedicated TDRs.
 
\subsection{  
Radiation levels on the LHCb detector for Upgrade II\label{subsec:level5.1}}
 
LHCb is and will remain a very lightweight detector from the IP up to the Electromagnetic Calorimeter. Any new choice in technology for Upgrade II in this area will have a similar distribution of mass as the current Upgrade I design. Therefore, preliminary values for radiation level estimators, e.g. TID, can be obtained by scaling results from Monte Carlo simulations of the current setup with the increase in luminosity, as only massive changes in material will significantly affect the radiation environment. Simulations of the LHCb radiation environment have been benchmarked using a variety of measurements, pointing to an accuracy within a factor of 2 or better within most of the detector volume for Run~1 conditions~\cite{Karacson:2243499}. Run~2 measurements currently under preliminary analysis indicate the same behavior.

However, the overall increase in luminosity and, as a consequence, higher radiation levels in the experimental cavern will be a serious concern in terms of radiation damage for certain sub-detectors. Operational and lifetime reliability will require careful analysis in the evaluation of final technological choices.
The radiation environment of ECAL in particular is going to be severely impacted due to its relatively heavy mass. The modules close to the beam line already exhibit a strong degradation in performance due to increased optical attenuation. A higher pile-up during Run 5 will cause a combination of overlapping particle showers leading to the accumulation of damage effects at MGy level and in parallel a reduction in resolution. In order to counter these effects, Upgrade II foresees a fundamental restructuring of the central ECAL volume. Enhancing the detector's sensitivity will require a smaller cell size in central regions as well as a different converter such as tungsten instead of the lead currently used. Fast timing information will also be crucial to match measurements to individual p--p collisions.
In turn, the envisaged remodeling of the central area of ECAL as well as the potential exchange of the whole volume of the Hadronic Calorimeter (HCAL) will have a significant influence on the radiation field in the LHCb detector. The introduction of tungsten in the innermost ECAL modules will change the characteristics of the particle backsplash. The induced effect on the readout electronics of the upstream close by SciFi and Mighty tracker as well as RICH2 will need to be evaluated.
In this regard, one also has to consider the impact of the resulting 1MeVn-eq fluence in particular with relation to photo-detectors, including those for the new nearby TORCH. Shielding solutions such as the large neutron shielding installed in Upgrade I to protect the SciFi silicon photo-multipliers (SiPMs) from ECAL backsplash, will have to be re-evaluated for Upgrade II and localized solutions fitting in the available space investigated. 
Protecting the cryogenics area in UX85 by installing a large shielding wall, as described in section~\ref{sec:level6}, is also expected to have a slight increase in both HEH fluence and 1MeVn-eq fluence on some of the sub-detectors in its vicinity. This may influence for example the dark current in the SciFi SiPMs which is directly related to 1MeVn-eq fluence, potentially affecting detector performance.
In addition to the direct impact of increased dose and fluence levels on the performance and life expectancy of the detector, increased residual dose levels will also impact maintenance operations during access. As an example, the added shielding material next to a frequented staircase close to the LHCb dipole on the cryogenics side of the experimental hall is expected to increase residual dose levels in the area compared to the currently installed fence. Dedicated studies to evaluate the implications of Upgrade II on radiation protection will be needed.

\section{Implications on the LHC electronic equipment\label{sec:level6}}
The various IR8 alcoves shown in Fig.~\ref{fig:geo_cavern} host several electronic systems, all potentially sensitive to radiation-induced damage. In particular, cryogenic  racks include elements such as Mecos Active Magnetic Bearings (AMB), which may lead to unwanted LHC downtime in the event of a radiation-induced failure.
After the Upgrade~II of LHCb, the foreseen annual luminosity of ${50\,\mathrm{fb^{-1}}}$ would increase radiation levels by a factor 7.5 compared to the entire Run~2, by more than a factor~20 compared to 2018 (which cumulated $2.5\,\mathrm{fb^{-1}}$, i.e. the maximum annual luminosity reached so far) and by more than a factor~5 compared to the maximum annual values expected after Upgrade I (Run~3--4). \\
  \begin{figure}[!h]
      \centering
      \begin{overpic}[trim=0.0cm 0.0cm 0.0cm 1.1cm, clip=true,width=1\columnwidth]{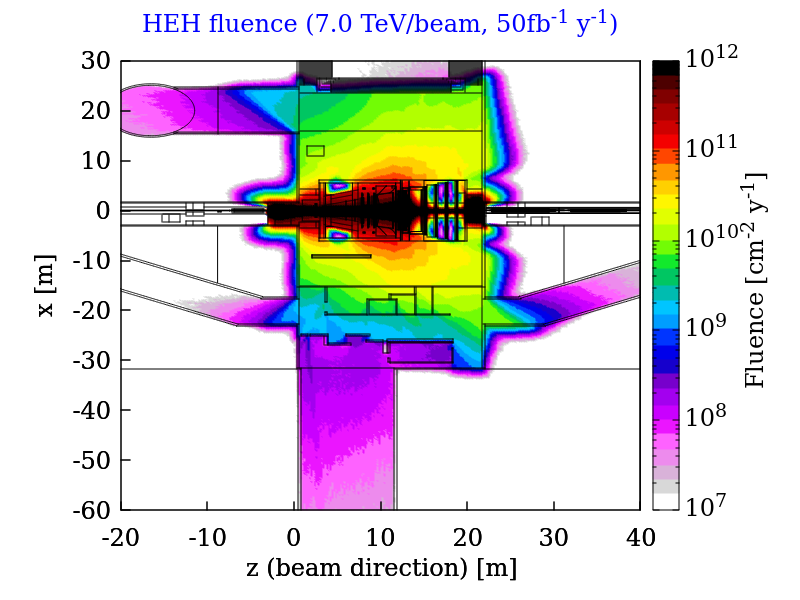}
        \put(430,565){\linethickness{0.3mm}\color{black}\vector(0.4,-0.5){30}}
        \put (400,570) {\textbf{\color{black} LHCb}}
      \put(250,420){\linethickness{0.3mm}\color{black}\vector(0.9,-0.05){160}}
        \put (170,430) {\textbf{\color{black} UX85}}
        \put (430,340) {\textbf{\color{white} US85}}
        \put(365,290){\linethickness{0.5mm}\color{white}\polygon(0,0)(245,0)(245,100)(0,100)}%
        \put (170,300) {\textbf{\color{black} UL84}}
        \put (680,300) {\textbf{\color{black} UL86}}        
        \put (370,150) {\textbf{\color{black} UW85}}    
        \end{overpic}
      \caption{Top view of annual HEH fluence (averaged in the third dimension between -1.1 m and 1.1 m with respect to the beam height) during Run~5, assuming the same IR8 layout as in Run~3-4 and an annual luminosity of ${50\,\mathrm{fb^{-1}}}$. The simulation assumed upward polarity of the LHCb spectrometer and an external horizontal crossing of $250~\mu \mathrm{rad}$, pointing inside the ring, at $\sqrt{s}=14\,\mathrm{TeV}$. 
      }
      \label{fig:run5_top}
    \end{figure}
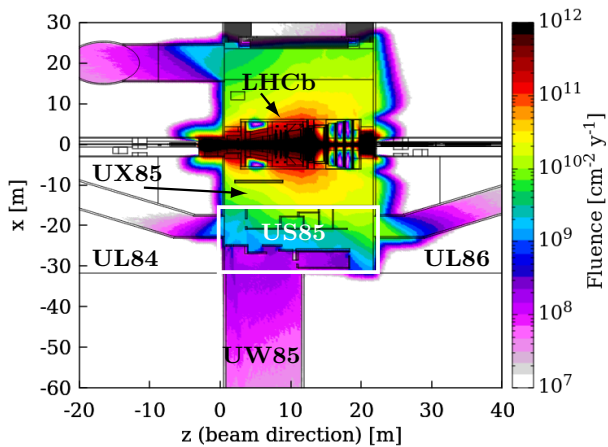 
The results shown in the following plots refer to external crossing on the horizontal plane and upward polarity of the LHCb spectrometer. However, given the proximity to the interaction point and the fact that we are interested here in high transverse momentum ($p_{\textrm T}$) debris, the crossing configuration has a negligible effect in view of the degree of accuracy required in these calculations. On the contrary, it has a major role in the assessment of radiation levels in the accelerator magnets, as discussed later in Section~\ref{sec:level7} and in~\cite{PhysRevAccelBeams.26.061002}.
Figure~\ref{fig:run5_top} shows the resulting annual HEH fluence distribution in the cavern and adjacent service areas. The conical propagation of the collision debris within UX85 (from left to right in Fig.~\ref{fig:run5_top}) extends to the nearby alcoves. This pattern originates from the asymmetric position of IP8 with respect to the centre of UX85 and the shape of the LHCb detector, as described in Section~\ref{sec:level3}. Radiation levels in the service tunnels (UL84 and UL86) are not symmetric, despite the symmetric layout of the galleries, being the electronics hosted in the UL86 tunnel significantly more exposed than the one in UL84. On the same premise, one observes relatively lower fluences on the left side of US85, near the UL84 entrance, and relatively higher fluences on the right side of US85, near the UL86 entrance. 

\begin{figure}[h!]
      \centering
        \begin{overpic}[trim=0.0cm 0.2cm 0.4cm 1.0cm, clip=true,width=1.0\columnwidth]{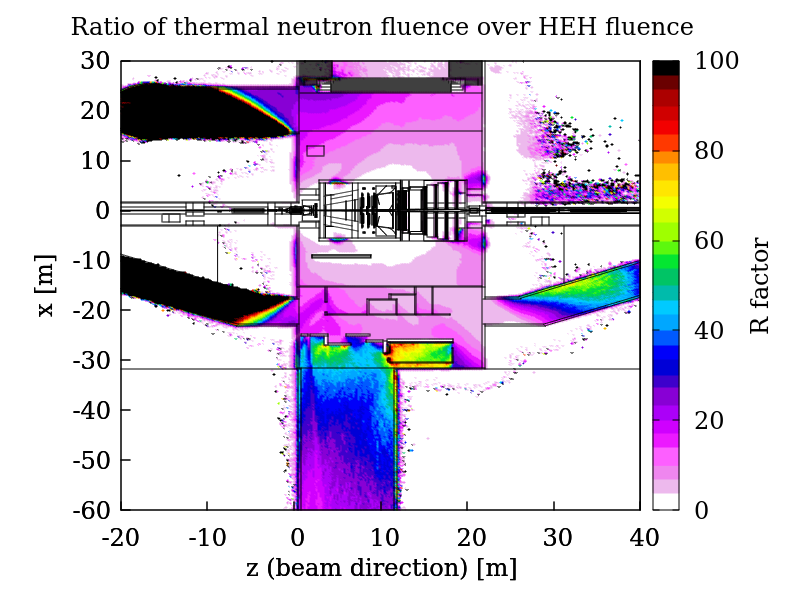}
      \put(430,565){\linethickness{0.3mm}\color{black}\vector(0.4,-0.5){30}}
      \put (400,570) {\textbf{\color{black} LHCb}}
      \put(250,420){\linethickness{0.3mm}\color{black}\vector(0.9,-0.05){160}}
        \put (170,430) {\textbf{\color{black} UX85}}
        \put (430,340) {\textbf{\color{white} US85}}
        \put(365,290){\linethickness{0.5mm}\color{white}\polygon(0,0)(245,0)(245,100)(0,100)}%
        \put (170,300) {\textbf{\color{black} UL84}}
        \put (680,300) {\textbf{\color{black} UL86}}        
        \put (370,150) {\textbf{\color{black} UW85}}    
        \end{overpic}
        \caption{Top view of the R-factor map at beam height.}
      \label{fig:Rfactor}
\end{figure}
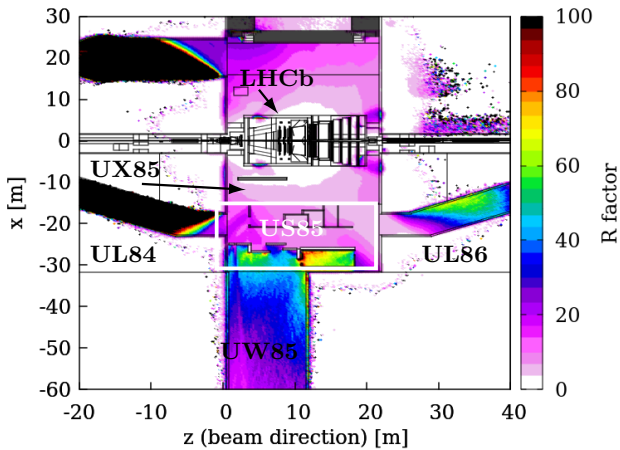

  \begin{table}
    \centering
    \caption{Annual HEH fluence values foreseen after the Upgrade II of LHCb in reference electronics positions within the indicated areas. Due to radiation level gradients, the levels cannot be generalized to the entire galleries.}
    \label{table:Run5_table}
    \begin{tabular}{ccc}
      \toprule
      Area & Level & HEH fluence\,$[\mathrm{cm^{-2}/\,50\,fb^{−1}}]$\\ 
      \toprule
	  \multirow{3}{*}{UX85} & 1 & {$10^{10}$} \\
	  & 0 & {$7\cdot10^{9}$} \\
	  & -1 & {$2\cdot10^{10}$} \\
	  \colrule
	  \multirow{3}{*}{US85} & 0 & {$2\cdot10^{9}$} \\
	  & 1 & {$2\cdot10^{9}$} \\
	  & 2 & {$7\cdot10^{9}$} \\
	  \colrule
	  {UL84} & 0 & {$4\cdot10^{8}$} \\
	  \colrule
	  {UL86} & 0 & {$10^{9}$} \\
    \botrule
    \end{tabular}
  \end{table} 
Table~\ref{table:Run5_table} summarizes the simulated HEH fluence values in the critical areas where electronic devices are placed. 
They are well above the earlier mentioned threshold for radiation safe areas. The HEH fluence in the safe room (see Fig.~\ref{fig:geo_cavern}), that is already equipped with shielding, ranges up to $3\cdot10^{8}\mathrm{cm^{-2}y^{-1}}$, as shown in Fig.~\ref{fig:run5_top} and in the left frame of Fig.~\ref{fig:run5_cross}. The accelerator equipment placed in UX85 is expected to be exposed to HEH fluences ranging from $7\cdot10^{9}\mathrm{cm^{-2}y^{-1}}$ 
to few $10^{10}\,\mathrm{cm^{-2}y^{-1}}$. 
A HEH fluence of $4\cdot10^{8}\mathrm{cm^{-2}y^{-1}}$ is calculated at the corner between US85 and UL84, where a rack hosting AMB Mecos units presently sits. Further along the UL84, the HEH fluence displays a high gradient and after few meters it has been evaluated to be around $10^{7}\,\mathrm{cm^{-2}y^{-1}}$ in correspondence with another electronics rack.
The HEH fluence of $10^{9}\,\mathrm{cm^{-2}y^{-1}}$ in UL86 has been estimated at about 5 metres from the beginning of the tunnel, close to an installed rack.
Moreover, the contribution of thermal neutrons, which can provoke soft errors in sensitive electronic systems, is dominant in the ULs as shown by the R-factor map illustrated in Fig.~\ref{fig:Rfactor}. The R-factor is defined as:
\begin{equation} \label{rfac}
R = \frac{\Phi_{n-th}}{\Phi_{HEH}}
\end{equation}
(see Section~\ref{sec:level3}) and indicates the relative importance of thermal neutrons with respect to HEH. It is much higher in the ULs and in the shielding areas like UW85 and the safe room in the US85, compared to the experimental cavern (around the detector and in UW85). In this case the R-factor was not evaluated in the LHC tunnel, where its typical value is around 1.5. 

In terms of SEE risk after Upgrade II, projections can be made based on the past operation of the presently installed electronic systems, as well as radiation tests carried out on selected elements (e.g., the AMB Mecos). The results do not rule out the risk of observing several SEEs per year, largely due to the expected increase of radiation levels caused by the higher luminosity. 

While local shields can be designed for thermal neutrons to mitigate soft SEE, extensive shielding solutions appear nonetheless to be necessary in order to mitigate the SEE risk due to high energy hadrons, if a relocation of the electronics racks is not possible.\\

\subsection{Benchmarking studies}

In addition to the Beam Loss Monitor (BLM) benchmark previously published~\cite{PhysRevAccelBeams.26.061002}, further comparisons between measurements and radiation levels predicted by the described FLUKA model around the LHCb detector were carried out for two experimental setups. 
One was obtained by comparing the Radiation Monitor (RadMon)~\cite{Spiezia:2011660} measurements collected during Run~2 (for an integrated luminosity of $6.6\,\mathrm{fb}^{-1}$) with the HEH fluence calculated at their positions in the experimental areas. 
In most cases the agreement is within a factor 2, while in few others the simulation overestimates the fluence by up to a factor of 2.6. This can be explained by the complexity of the cavern geometry and the typical underestimation of the actual material budget due to neglected elements.

Another benchmarking study was done for the experimental area UX85A, as indicated in Fig.~\ref{fig:geo_cavern}, on the opposite side with respect to the alcoves discussed in the previous section. A concrete wall more than three metres thick protects this area, which allowed a computing farm to be installed close to the detector. Due to its refurbishment to comply with the Upgrade I online computing needs, the farm was relocated on the surface of Point 8 during LS2. Currently the structure immediately behind this shielding wall still contains detector control equipment. This area was considered for hosting the CODEX experiment in the future\cite{Aielli_2020, dey2019background}. During 2022, HEH, n-th and muon flux measurements in UX85A were performed in the candidate position of the CODEX experiment. In particular, $\Phi_{HEH}$ and $\Phi_{n-th}$ measurements were performed at beam height with the Battery powered version (BatMon)~\cite{Zimmaro:2823933} of the RadMon and turned out to be below the detection threshold, in agreement with the simulated neutron fluence barely exceeding $10^6\,\mathrm{cm}^{-2}$ over the considered period. At the same time, The muon flux measurement obtained with an R2E BLM based on the Timepix3~\cite{Poikela2014} technology, namely $2.35\times10^{-3}\,\mathrm{counts}\,\mathrm{cm}^{-2}\,\mathrm{s}^{-1}$, matched very well the FLUKA prediction.

\subsection{Shielding design and engineering constraints }
 The study of radiation levels in the experimental areas revealed the need to protect all electronic devices to prevent the occurrence of SEEs. Given the scale of the areas to be protected, namely three levels in height over the full length of the cavern, the design of such a shielding is shaping up to be the largest ever made at CERN.

The shielding has to be integrated in UX85 between the detector and the cryostats, aiming to decrease the radiation levels to values as after the Upgrade I. As the annual luminosity increase between Upgrade I and Upgrade II amounts to a factor 5, the shielding should guarantee that at least $80\%$ of the hadronic population undergoes an inelastic nuclear interaction through it. The probability that a particle has an inelastic nuclear interaction along a travelled distance $l$ inside a material is given by the cumulative distribution function: 
\begin{eqnarray}
P(l)= \int_{0}^{l}p(l')\,dl'=\,1-e^{-\frac{l}{\lambda_I}}
\label{eq:one}.
\end{eqnarray}
where $\lambda_{I}$ is the so-called inelastic scattering length (whose dependence on the particle energy can be neglected above few hundred MeV).
Eq.~(\ref{eq:one}) implies that 63\% of the hadron population undergoes an inelastic nuclear collision through a path length $\lambda_I$ and 86\% within $2\lambda_I$.
For high energy neutrons in concrete, $\lambda_{I}$ is about $40\,\mathrm{cm}$~\cite{Workman:2022ynf}. Therefore, a 80 cm thick concrete wall is expected to assure the required attenuation. Such a thickness cannot be guaranteed for the entire structure, due to the presence of cryostats and structural elements. As a solution, iron blocks replace the concrete ones. In fact, $\lambda_{I}$ for iron is about 17 cm, allowing for the replacement of 80~cm concrete blocks with standard 40~cm iron blocks. The considered wall does not completely seal off the LHCb detector from the adjacent areas because, despite its 11-metre height, it does not reach up to the ceiling and, moreover, small apertures for cable passage and accessibility to the cryogenic instrumentation cannot be avoided.
\begin{figure}[!ht]
      \centering
      \includegraphics[width=0.8\columnwidth]{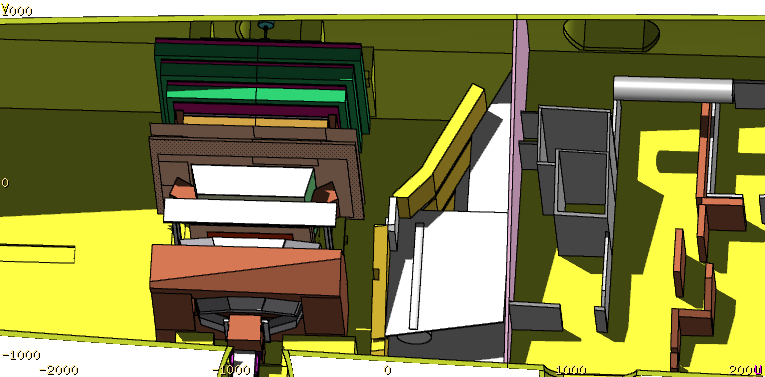}
      \caption{3D FLUKA geometry of UX85 and US85, including the LHCb detector and the shielding wall.}
      \label{fig:wall3Dgeo}
    \end{figure} 
  \begin{figure}[!ht]
      \centering
      \includegraphics[width=0.8\columnwidth]{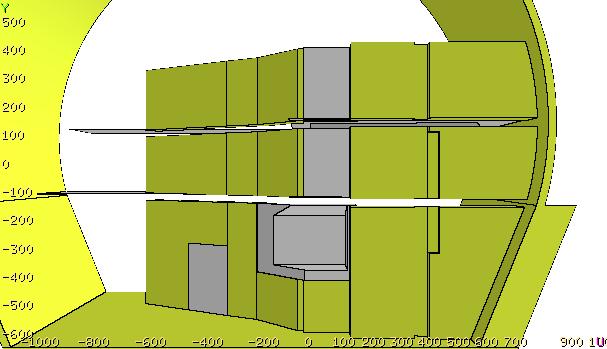}
      \caption{3D FLUKA geometry of the shielding wall seen from the LHCb experiment side.}
      \label{fig:wall3D}
    \end{figure} 

\begin{figure}[h!]
      \centering
        \begin{overpic}[trim=0.0cm 0.0cm 0.0cm 1.1cm, clip=true,width=1\columnwidth]{  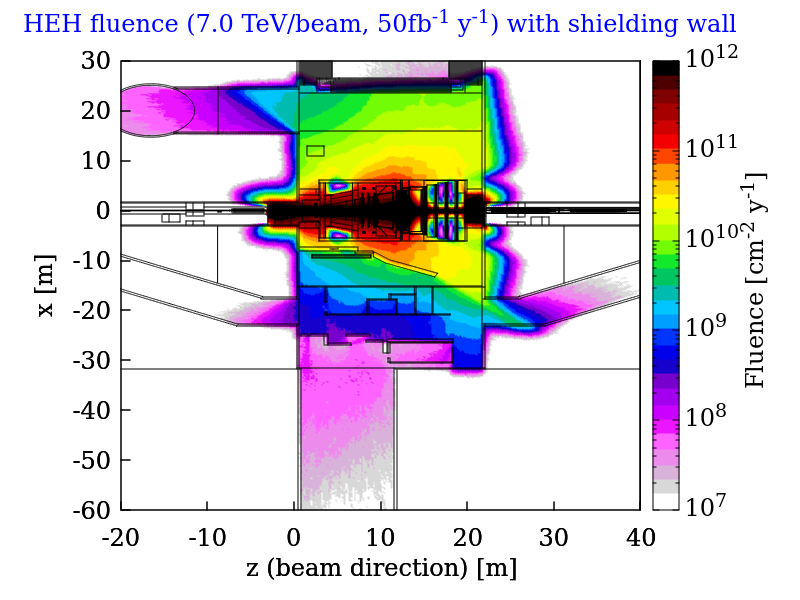}
        \put(430,565){\linethickness{0.3mm}\color{black}\vector(0.4,-0.5){30}}
        \put (400,570) {\textbf{\color{black} LHCb}}
        \put(700,525){\linethickness{0.3mm}\color{red}\vector(-0.9,-0.5){200}}
        \put (640,530) {\textbf{\color{red} Sh.wall}}        \put(250,420){\linethickness{0.3mm}\color{black}\vector(0.9,-0.05){160}}
        \put (170,430) {\textbf{\color{black} UX85}}
        \put (430,340) {\textbf{\color{white} US85}}
        \put(365,290){\linethickness{0.5mm}\color{white}\polygon(0,0)(245,0)(245,100)(0,100)}%
        \put (170,300) {\textbf{\color{black} UL84}}
        \put (680,300) {\textbf{\color{black} UL86}}        
        \put (370,150) {\textbf{\color{black} UW85}}    
        \end{overpic}
      \caption{Top view of annual HEH fluence (averaged in the third dimension between -1.1  m  and  1.1  m  with respect to  the  beam  height) during Run~5 in the presence of the shielding wall. The same simulation conditions as in Fig.~\ref{fig:run5_top} apply.}
      \label{fig:run5_top_wall}
    \end{figure}

The effect of the wall, displayed in Figures~\ref{fig:wall3Dgeo} and \ref{fig:wall3D}, can be appreciated in Fig.~\ref{fig:run5_top_wall}. The latter should be compared with the corresponding plot in Fig.~\ref{fig:run5_top} showing the HEH fluence without the wall. 

\begin{figure*}
        \begin{overpic}[trim=0.0cm 0.2cm 0.4cm 1.0cm, clip=true,width=0.45\linewidth]{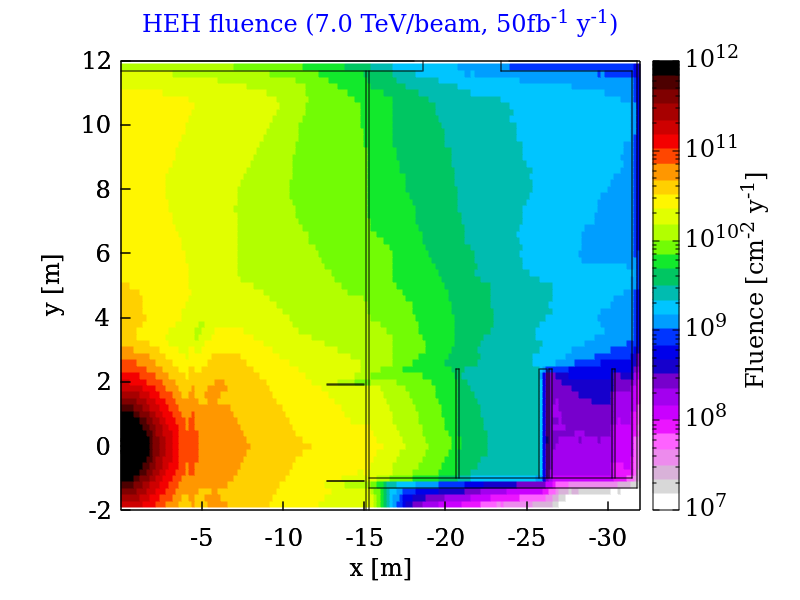}
        \put (155,370) {\textbf{\color{black} LHCb}}
        \put (320,430) {\textbf{\color{black} UX85}}
        \put (600,530) {\textbf{\color{black} US85}}
        \put(480,460){\linethickness{0.5mm}\line(1,0){300}}%
        \put(480,280){\linethickness{0.5mm}\line(1,0){300}}%
        \put (600,340) {\textbf{\color{black} Safe~room}}
        \put(750,330){\linethickness{0.3mm}\color{black}\vector(0,-1){100}}
        \end{overpic}
        \begin{overpic}[trim=0.0cm 0.2cm 0.4cm 1.0cm, clip=true,width=0.45\linewidth]{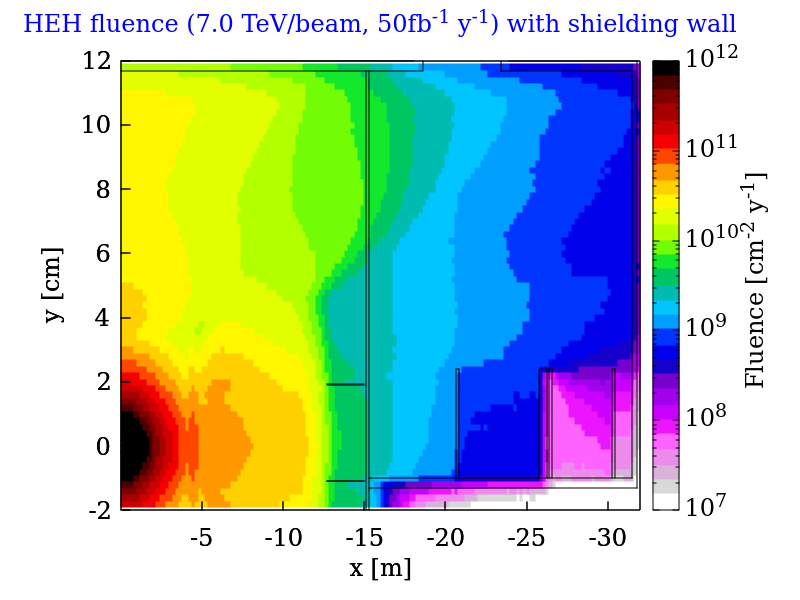}
        \put (155,370) {\textbf{\color{black} LHCb}}
        \put (320,430) {\textbf{\color{black} UX85}}
        \put (600,530) {\textbf{\color{black} US85}}
       \put(400,100){\linethickness{1.5mm}\color{red}\line(0,1){280}}%
        \put(480,460){\linethickness{0.5mm}\line(1,0){300}}%
        \put(480,280){\linethickness{0.5mm}\line(1,0){300}}%
        \put (600,340) {\textbf{\color{black} Safe~room}}
        \put(750,330){\linethickness{0.3mm}\color{black}\vector(0,-1){100}}
        \put (155,520) {\textbf{\color{red} Shielding~wall}}
       \put(250,510){\linethickness{0.3mm}\color{red}\vector(0.6,-1){100}}
        \end{overpic}
      \caption{Cross-sectional view of annual HEH fluence (averaged along the beam direction between 13 m and 17 m from IP8) after the Upgrade II of LHCb with the current layout (left plot) and with the implementation of the shielding wall (right plot). The same simulation conditions as in Fig.~\ref{fig:run5_top} apply.}
 \label{fig:run5_cross}
\end{figure*}
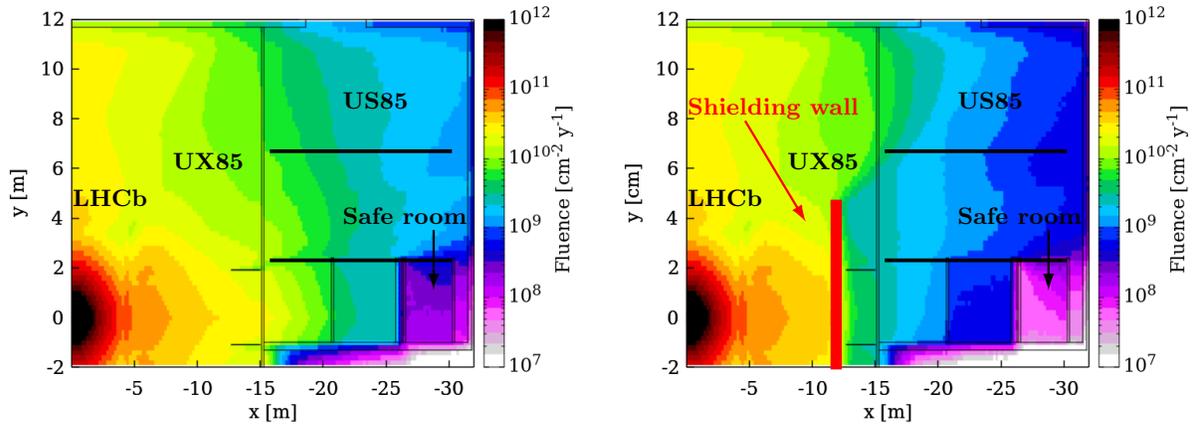 


The shielding wall in UX85 provides a reduction of the HEH fluence by almost a factor 5 at the UL84-US85 corner ($\mathrm{z}=0$) and about a factor 3 when entering the UL84 ($\mathrm{z}\leq2\,\mathrm{m}$).

On the other side, the situation is different at the boundary between US85 and UL86, since the shielding has been designed according to the shape of the detector and no protection effect is achieved there.\\
 

Figure~\ref{fig:run5_cross} shows the picture at the various levels of UX85 and US85 in the current layout and in the presence of the shielding, highlighting the contribution coming from the upper floors.
As for thermal neutrons, no adequate reduction can be obtained in US85, UL84 and UL86 and local shielding shall be designed to protect the racks in this regard.
Conversely, the reduction in terms of TID is even better than for HEH fluence, namely of about one order of magnitude in UX85 just behind the wall, from ~25 Gy without shielding down to a few Gy (per ${50\,\mathrm{fb^{-1}}}$).


\section{Implications on the LHC magnets\label{sec:level7}}
Concerning the implications of the Upgrade II of LHCb on the LHC magnets, the critical points are related to the absence of protection elements that are installed in IR1 and IR5 but considered unnecessary in IR8 up to the Run~3 luminosity. They are the absorber (TAS) protecting the first final focus quadrupole (Q1) and the physics debris collimators (TCL) in the matching section. On the other hand, an absorber (TANB) protecting the recombination dipole (D2) was already installed during LS2 in view of the current luminosity increase, and here we demonstrate its effectiveness even for the further upgrade of LHCb. Moreover, the superconducting separation dipole (D1), which was replaced by a normal-conducting version for IR1 and IR5, calls for a dedicated study in view of the Upgrade II luminosity specifications. We will describe the various points starting from the elements closest to IP8 (i.e., the conventional compensator magnets) and moving toward the arc.

\subsection{Normal-conducting compensator magnets}
The compensator magnets are the closest elements to the IP and thus, they are significantly exposed to radiation damage, especially their coils on the IP side. 
Since they are normal-conducting magnets, we are interested only in evaluating the TID level in the long term that could jeopardize the reach of the desired lifetime in terms of integrated luminosity. 
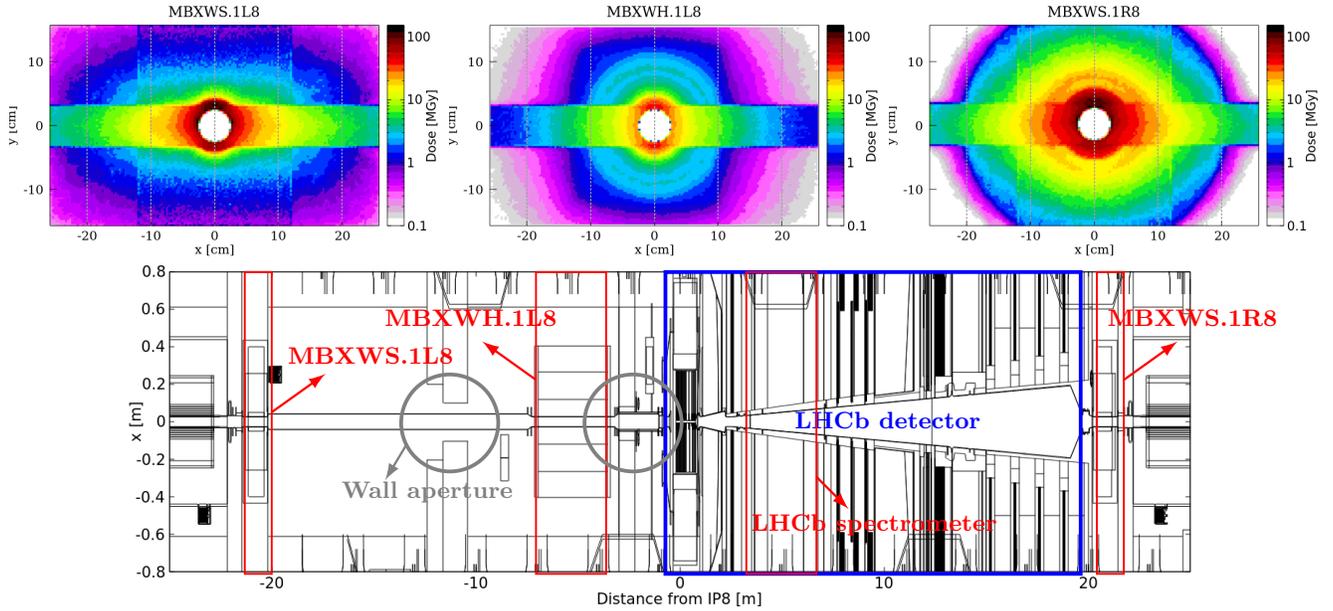
\begin{figure*}[!ht]
    \centering
    \includegraphics*[trim=1.5cm 3.5cm 0.5cm 0.0cm, clip=true,width=0.32\textwidth]{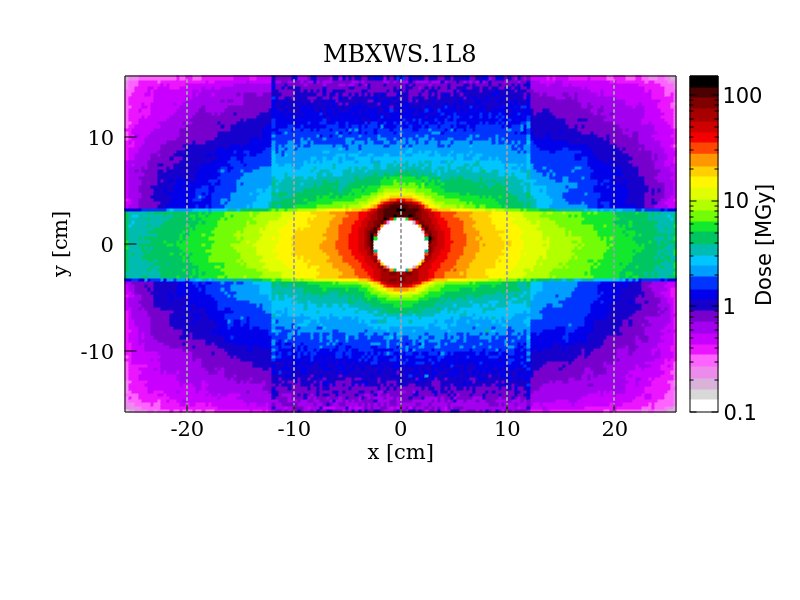}
    \includegraphics*[trim=1.5cm 3.5cm 0.5cm 0.0cm, clip=true,width=0.32\textwidth]{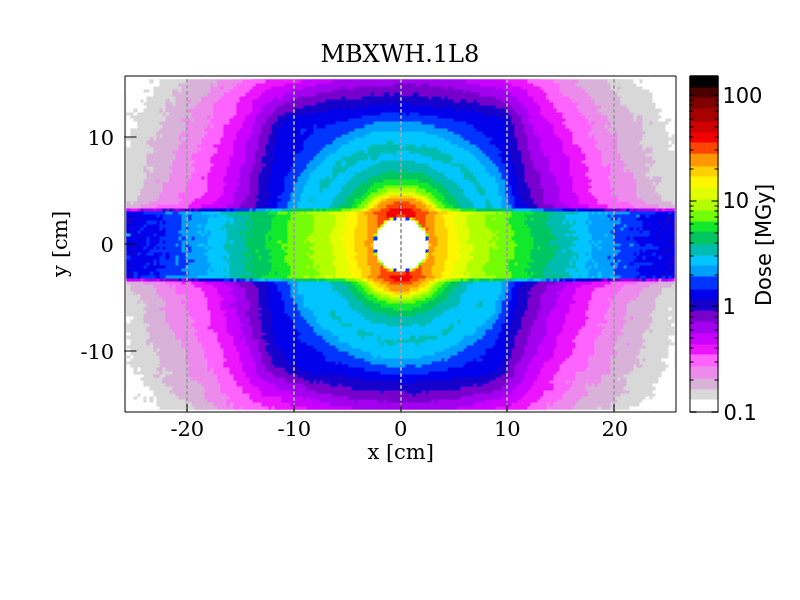}
    \includegraphics*[trim=1.5cm 3.5cm 0.5cm 0.0cm, clip=true,width=0.32\textwidth]{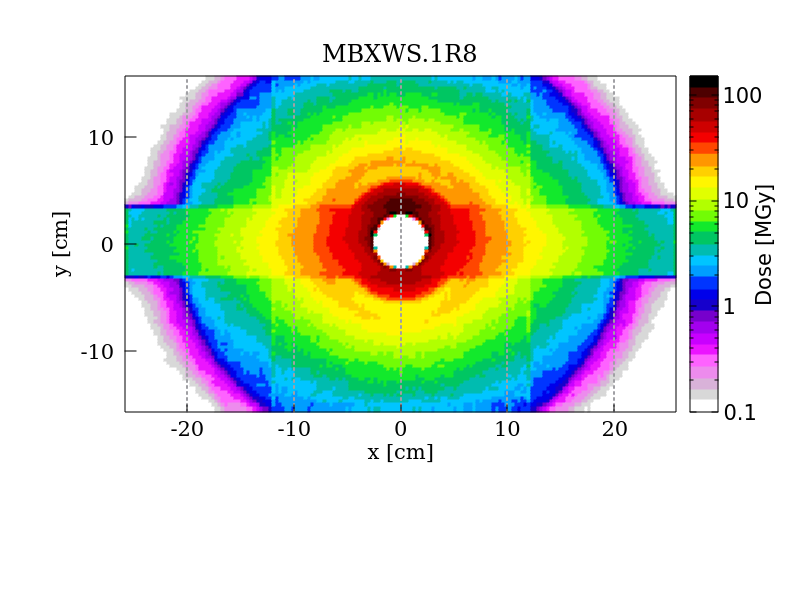}
       \begin{overpic}[trim=0.0cm 0.0cm 0.0cm 0.0cm, clip=true,width=0.8\textwidth]{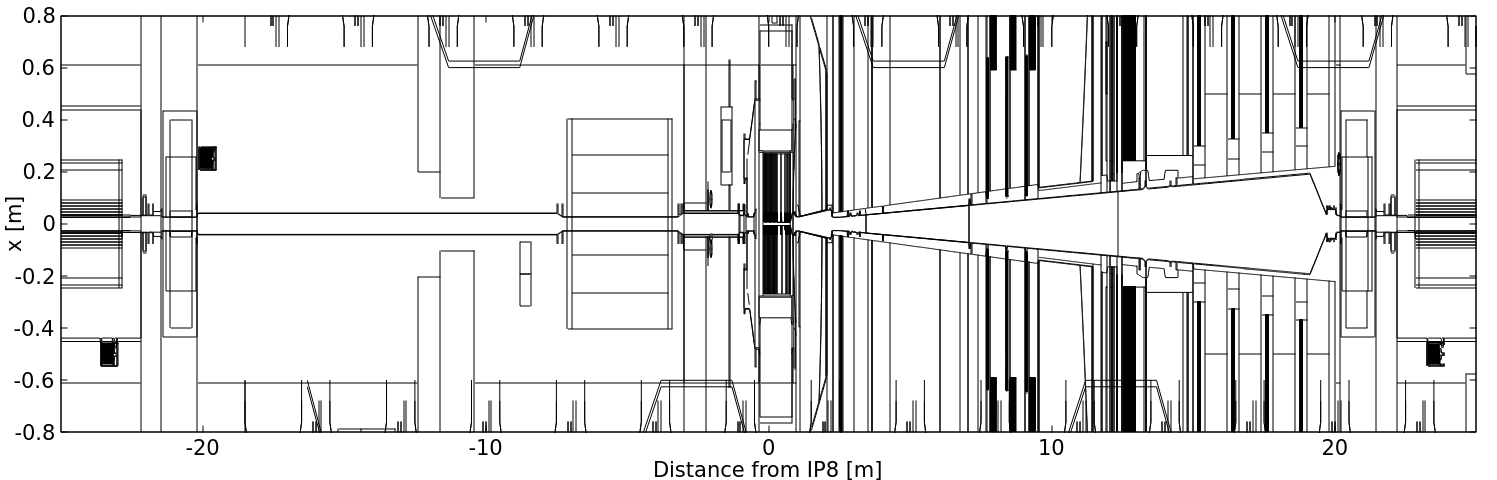}
        \put(110,30){\linethickness{0.25mm}\color{red}\polygon(0,0)(25,0)(25,280)(0,280)}%
        \put(135,180){\linethickness{0.3mm}\color{red}\vector(
        0.7,0.5){50}}
        \put (150,225) {\textbf{\color{red} MBXWS.1L8}}
        \put(500,30){\linethickness{0.5mm}\color{blue}\polygon(0,0)(385,0)(385,280)(0,280)}%
        \put (620,165) {\textbf{\color{blue} LHCb detector}}
        \put(575,30){\linethickness{0.25mm}\color{red}\polygon(0,0)(65,0)(65,280)(0,280)}%
        \put(640,120){\linethickness{0.3mm}\color{red}\vector(
        0.5,-0.5){30}}
        \put (580,70) {\textbf{\color{red} LHCb spectrometer}}
        \put(380,30){\linethickness{0.25mm}\color{red}\polygon(0,0)(65,0)(65,280)(0,280)}%
        \put(380,210){\linethickness{0.3mm}\color{red}\vector(
        -0.7,0.5){50}}
        \put (240,260) {\textbf{\color{red} MBXWH.1L8}}
        \put(900,30){\linethickness{0.25mm}\color{red}\polygon(0,0)(25,0)(25,280)(0,280)}%
        \put(925,210){\linethickness{0.3mm}\color{red}\vector(
        0.7,0.5){50}}
        \put (910,260) {\textbf{\color{red} MBXWS.1R8}}
        \put(200,100){\textbf{\color{gray}Wall aperture}}%
        \put(300,170){\linethickness{0.5mm}\color{gray}\circle{90}}%
        \put(470,170){\linethickness{0.5mm}\color{gray}\circle{90}}%
        \put(260,150){\linethickness{0.3mm}\color{gray}\vector( -0.3,-0.5){20}}
        \end{overpic}
    \caption{Transverse dose distribution at the maximum of the IP-face front coils of the indicated conventional compensator magnets. 
    External upward vertical crossing is assumed and values are normalized to $360 \,\mathrm{fb}^{-1}$. The bottom plot shows the geometry apertures around IP8, highlighting the location of the compensator magnets and the LHCb spectrometer in red and the shielding walls in grey.}
    \label{fig:2D_front_coils}
\end{figure*}
Figure~\ref{fig:2D_front_coils} shows the transverse dose distribution on the IP-side front coil for the three magnets placed at the positions indicated in the geometry sketch in the bottom plot. The left and right top plots exhibit different distributions at the same distance from IP8 due to the non-symmetrical layout. The compensator magnets on the left (MBXWS.1L8 and MBXWH.1L8) are partially protected by concrete walls intercepting a fraction of the collision debris but the particles at lower $p_T$ travelling through the aperture restrictions. The two normal-conducting dipoles have the same aperture and the long one (MBXWH), being closer to IP8, is impacted by particles travelling at a larger angle and therefore carrying a lower energy, which results in a lower dose. On the right side, the magnet is just downstream of the LHCb detector. The most impacted part is the edge between the upper coil and the beam pipe, because of the upward external vertical crossing assumed in the simulation. The longitudinal peak is deep, close to the magnet yoke as in Fig.~\ref{fig:dose_MBXWS_longitudinal} (left frame).
By the end of Run~6, assuming no Upgrade II, the expected peak dose is around $45\,\mathrm{MGy}$ for each of the short compensator magnets (MBXWS.1L8 and MBXWS.1R8) and around $15\,\mathrm{MGy}$ for the long one, for fixed external angle polarity. Otherwise, the simulated peak exceeds $120\,\mathrm{MGy}$ for each of the short compensator magnets (see Fig.~\ref{fig:dose_MBXWS}), well above the estimated damage limit~\cite{1952741}.
The regular inversion of the polarity of the external vertical crossing could decrease the dose to around $100\,\mathrm{MGy}$, which is still too high. In contrast, even in the case of Upgrade II, no problems are expected for the MBXWH coils exposed to a dose of $40\,\mathrm{MGy}$. A mitigation strategy has then to be applied to protect the MBXWS coils on both sides of IP8.

\subsubsection{Front face shielding}
The proposed solution to reduce the peak dose is the placement of a tungsten shielding close to the vacuum chamber and almost in contact with the coil. 
\begin{figure}[!tbh]
    \centering   
    \includegraphics*[trim=5.0cm 3.0cm 12.0cm 3.0cm, clip=true,width=0.45\columnwidth]{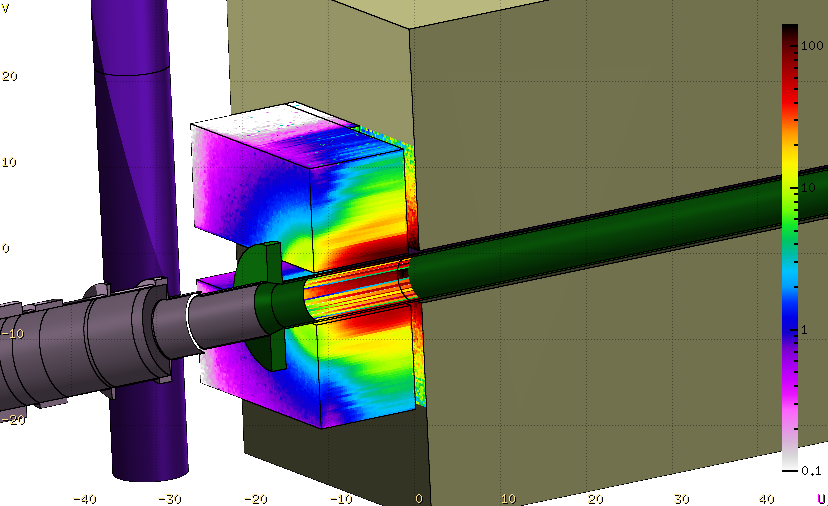}
    \includegraphics*[trim=5.0cm 3.0cm 12.0cm 3.0cm, clip=true,width=0.45\columnwidth]{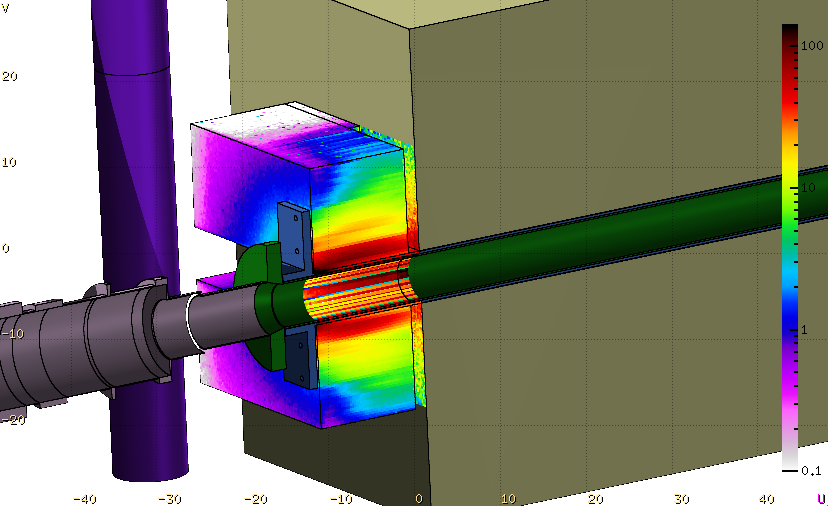}
    \includegraphics*[angle=-90, trim=0.0cm 0.0cm 0.0cm 0.0cm, clip=true,width=0.90\columnwidth ]{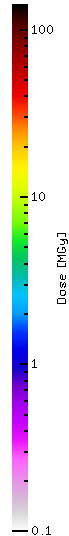}
    \caption{Three-dimensional dose maps plotted on a longitudinal cutaway of the right short compensator magnet for the current geometry (left) and with the addition of a tungsten shielding with a thickness of 4 cm (right). Values are normalized to $360\,\mathrm{fb^{-1}}$. Upward external vertical crossing is assumed. The scoring mesh is Cartesian with transverse and longitudinal resolution of 3 mm.
    }
    \label{fig:dose_MBXWS_longitudinal}
\end{figure}
\begin{figure}[!h]
    \centering   
    \includegraphics*[trim=1.5cm 2.5cm 2.5cm 0.0cm, clip=true,width=0.95\columnwidth]{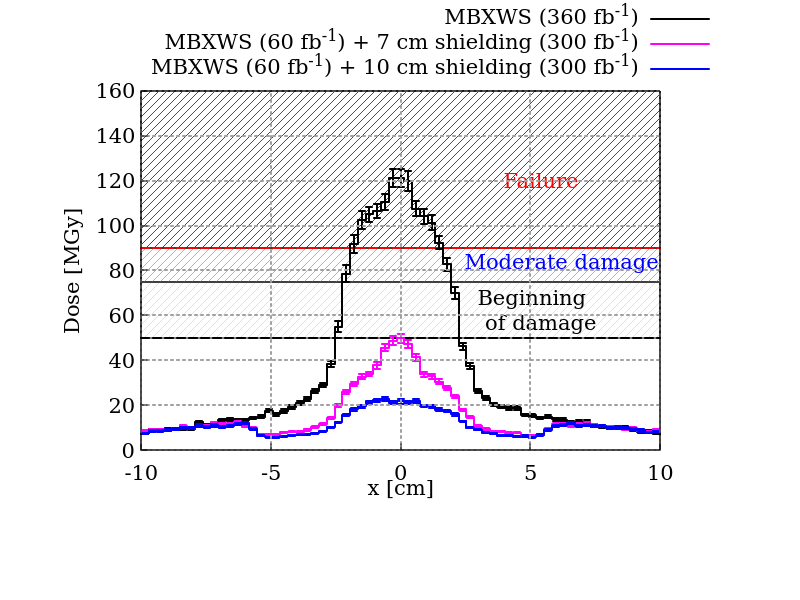}
    \caption{Transverse peak dose profiles on the front upper coil of the MBXWS on the left side of IP8, evaluated with a longitudinal resolution of 5 cm for different shielding options, implemented (if any) after accumulating $60\,\mathrm{fb}^{-1}$ during Run~3 and Run~4. Upward external crossing and downward polarity of the LHCb spectrometer have been assumed.}
    \label{fig:dose_MBXWS}
\end{figure}
The coils are mainly impacted by charged pions with energy of several hundred GeV and photons (from neutral pion decay) with a spectrum peaked around 250 GeV~\cite{PhysRevAccelBeams.26.061002}. The dominant contribution to local dose deposition is due to the electromagnetic cascade, developed by high energy photons through electron--positron pair production and in turn electron/positron Bremsstrahlung generating new photons. The frequency of these processes is linked to the radiation length $X_{0}$ of the material, which is equal to 7/9 of the photon mean free path for pair production and is also the average distance to reduce the electron/positron energy by a factor of 1/e.
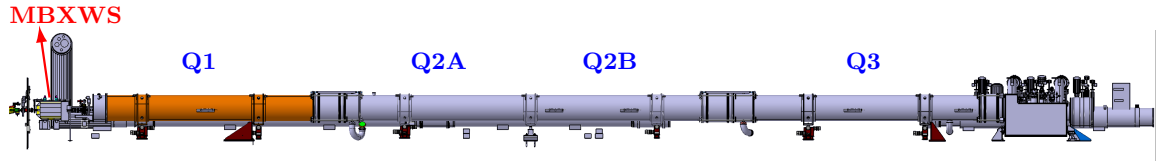
\begin{figure*}[!th]
    \centering
        \begin{overpic}[trim=0.0cm 0.0cm 0.0cm 0.0cm, clip=true,width=0.85\textwidth]{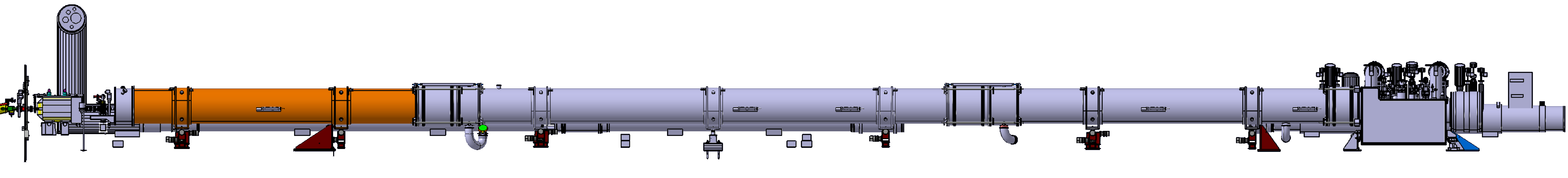}
        \put(35,55){\linethickness{0.3mm}\color{red}\vector( -0.1,0.7){9}}
        \put (0,120) {\textbf{\color{red} MBXWS}}

        \put (150,80) {\textbf{\color{blue} Q1}}
        \put (350,80) {\textbf{\color{blue} Q2A}}
        \put (500,80) {\textbf{\color{blue} Q2B}}
        \put (730,80) {\textbf{\color{blue} Q3}}
        \end{overpic}
    \caption{Layout of the triplet on the right side of IP8.}
    \label{fig:Layout_R}
\end{figure*}
The geometrical increase of the number of shower particles comes to an end when the particle energy is reduced to the critical energy $E_{c}$ of the material, below which atom ionization and excitation become dominant with respect to radiative processes. Therefore, the shower maximum is approximately reached after a distance
\begin{eqnarray}
l_{max}= log(\frac{E_{0}}{E_{c}})+0.5
\label{eq:photon_cascade}.
\end{eqnarray}
in units of radiation length, where $E_{0}$ is the initial photon energy. 
For copper, $X_{0}=1.4\,\mathrm{cm}$ and $E_{c}=19\,\mathrm{MeV}$, which in our case yields the dose peak in the coils after more than 10 cm, as in the left frame of Fig.~\ref{fig:dose_MBXWS_longitudinal}.

By adding a few cm of an absorbent high-Z material upstream of the coils, such as tungsten ($X_{0}=3.5\,\mathrm{cm}$ and $E_{c}=8\,\mathrm{MeV}$), one can achieve the shift of the peak inside the shielding, as shown in the right frame of Fig.~\ref{fig:dose_MBXWS_longitudinal}. 

Assuming that this mitigation strategy will be put in place at the same time as the Upgrade II of LHCb (during LS4 preceding Run~5), Fig.~\ref{fig:dose_MBXWS} shows the effect of a tungsten shielding of 7 cm and 10 cm length, respectively, on the simulated dose for the short compensator magnet.
Even adding the contribution from the luminosity collected so far (around $10\,\mathrm{fb}^{-1}$ up to the end of 2022), mainly with external horizontal crossing that is less severe in regard to the compensator magnets, it appears that a shielding length of 7 cm can assure the preservation of the magnet functionality. 

\subsection{Final focus quadrupoles and separation dipole}

\begin{figure}[!tbh]
    \centering   
    \includegraphics*[trim=0.0cm 2.5cm 2.5cm 0.0cm, clip=true,width=0.95\columnwidth]{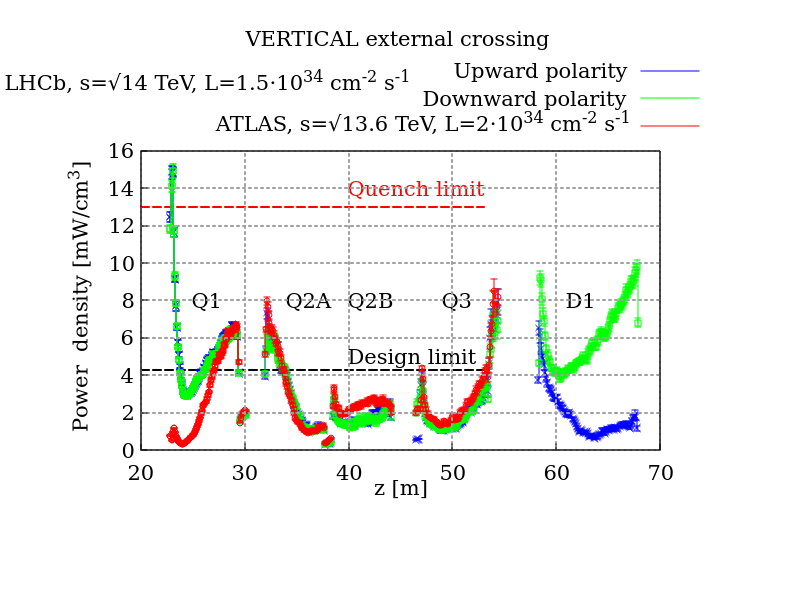}
    \caption{Longitudinal profile of peak power density in the superconducting coils along the triplet and the D1 on the right side of IP8 (IP1 for the red curve), which is at z=0. Values are averaged over the cable radial thickness and the azimuthal resolution is of $2^{\circ}$. Two curves, normalized to $1.5\cdot10^{34}\,\mathrm{cm}^{-2}\,\mathrm{s}^{-1}$, refer to the LHCb Upgrade II operational scenario with $\sqrt{s}=14\,\mathrm{TeV}$, external vertical crossing, and either upward (blue crosses) and downward (green squares) polarity of the LHCb spectrometer. The red curve refers to the ATLAS insertion (where no superconducting D1 is present) and an operational Run~3 scenario with $\sqrt{s}=13.6\,\mathrm{TeV}$ at $2\cdot10^{34}\,\mathrm{cm}^{-2}\,\mathrm{s}^{-1}$. The red dashed line represents the quench limit for the triplet quadrupoles and the black dashed line represents the design limit~\cite{Mokhov:613167}.}
    \label{fig:power_Q1-3_D1}
\end{figure}

The foreseen luminosity increase gives the need to evaluate the risk of quench of the superconducting quadrupoles of the final focus triplet (Q1--Q3), whose layout is shown in Fig.~\ref{fig:Layout_R}, and the subsequent separation dipole.
To this purpose, the longitudinal profile of peak power density in the inner superconducting coils along the triplet and the D1 is shown in Fig.~\ref{fig:power_Q1-3_D1}.
The maximum on the IP side of Q1 is higher than the quench limit, due to the absence of the TAS. In this regard, Fig.~\ref{fig:power_Q1-3_D1} highlights the effect of the TAS in protecting the first quadrupole of the ATLAS insertion. 
The quench limit for the triplet quadrupoles was estimated to be $13\,\mathrm{mW/cm^{3}}$~\cite{Mokhov:613167}, while the design limit was defined assuming a safety factor of three. In 2022, a luminosity record was set in ATLAS and CMS at more than twice their design luminosity of $10^{34}\,\mathrm{cm}^{-2}\,\mathrm{s}^{-1}$. No quenches were recorded in the triplet quadrupoles, which indicates that corresponding power density values around $7-8\,\mathrm{mW/cm^{3}}$ can be safely approached. Therefore, from our study one can conclude that, apart from the aforementioned concern for Q1, all other triplet quadrupoles could sustain the post-LS4 target instantaneous luminosity. However, Fig.~\ref{fig:power_Q1-3_D1} raises a second concern related to D1, which results significantly exposed on its non-IP side for one of the two spectrometer polarities (downward polarity). Moreover, another maximum on its IP side is present for both polarities of the LHCb spectrometer. In the absence of a precise assessment of the D1 quench limit, a possible mitigation measure was devised, as discussed in the following.

\begin{figure*}[!ht]
    \centering   
    \includegraphics*[trim=1.5cm 2.5cm 2.5cm 3.5cm, clip=true,width=0.95\textwidth]{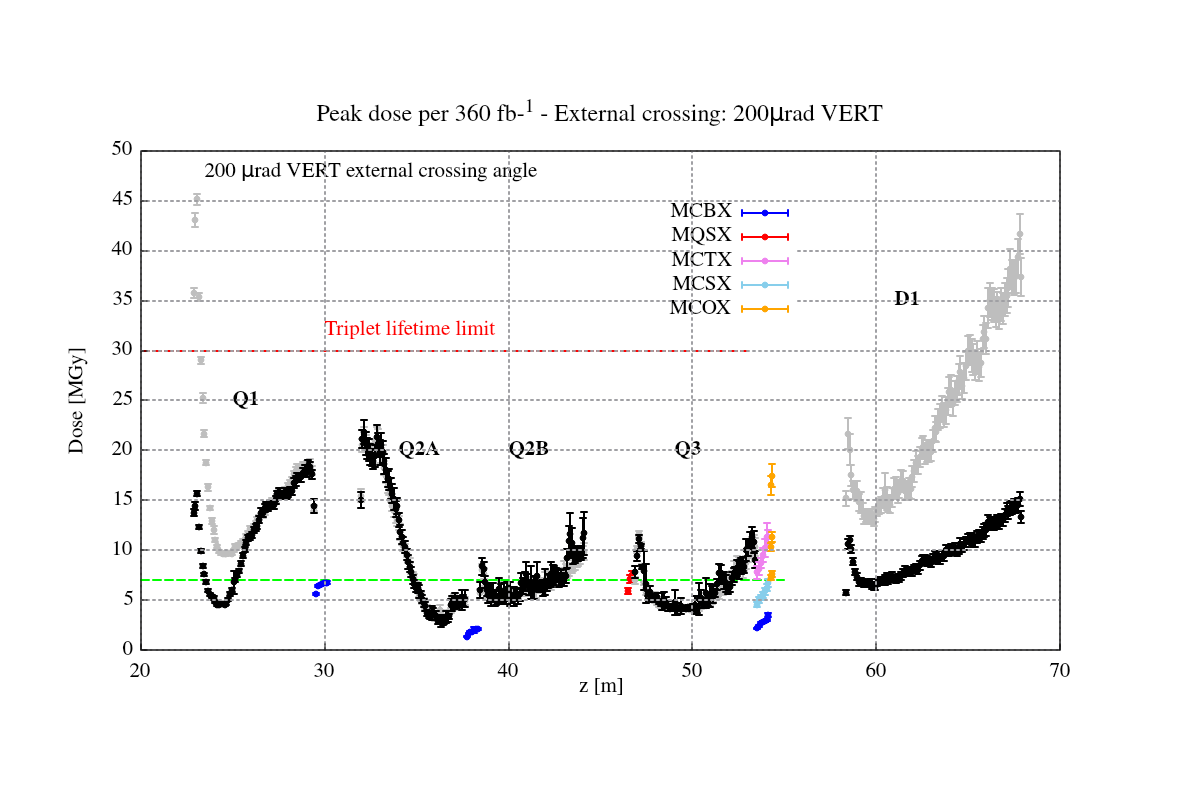}
    \caption{Longitudinal profile of peak dose in the superconducting coils along the triplet and the D1 on the right side with respect to IP8 (at z=0). The azimuthal and radial resolution is of $2^{\circ}$ and {3 mm}, respectively. External vertical crossing has been considered for $\sqrt{s}=14\,\mathrm{TeV}$. An integrated luminosity of $360\,\mathrm{fb}^{-1}$ is assumed to be collected half with either polarity of the LHCb spectrometer. The grey curve is for the current layout, while the black one combines $300\,\mathrm{fb}^{-1}$ with the shielding options discussed in Secs.~\ref{TAS-like} and \ref{D1} and $60\,\mathrm{fb}^{-1}$ without any mitigation measure up to the end of Run~4. The red dashed line represents the design lifetime of Q1-Q3~\cite{Tavlet} and D1, while the green one represents the design lifetime of the correctors~\cite{Corrector_limit}.}
    \label{fig:dose_Q1-3_D1} 
\end{figure*}

On the other hand, the dose distribution in the superconducting coils highlights an analogous problem for Q1 with regard to the damage threshold of the insulator, limiting the magnet lifetime. The peak dose profile in Fig.~\ref{fig:dose_Q1-3_D1} shows a maximum of {45 MGy} on the Q1 IP-face that is well above the level assumed for the triplet quadrupoles~\cite{Tavlet} (a small contribution from the integrated luminosity collected up to the end of 2022 should be added, amounting to less than {1.5 MGy} to the Q1 IP-face peak). 
As a consequence, a protection solution is required not only to avoid quench, but also to reach the desired integrated luminosity.
Furthermore, Fig.~\ref{fig:dose_Q1-3_D1} shows also a maximum of about 40 MGy on the D1 non IP-face, confirming the need for an additional D1 protection. In contrast to the peak power density profile, the local maximum on the D1 IP-side decreases to values below the limit because it is located on opposite azimuthal positions as a function of ther polarity of the LHCb spectrometer. 

\subsubsection{TAS-like absorber design \label{TAS-like}}
\begin{figure}[!tbh]
    \centering   
    \includegraphics*[trim=2.0cm 0.0cm 2.0cm 0.0cm, clip=true,width=0.45\columnwidth]{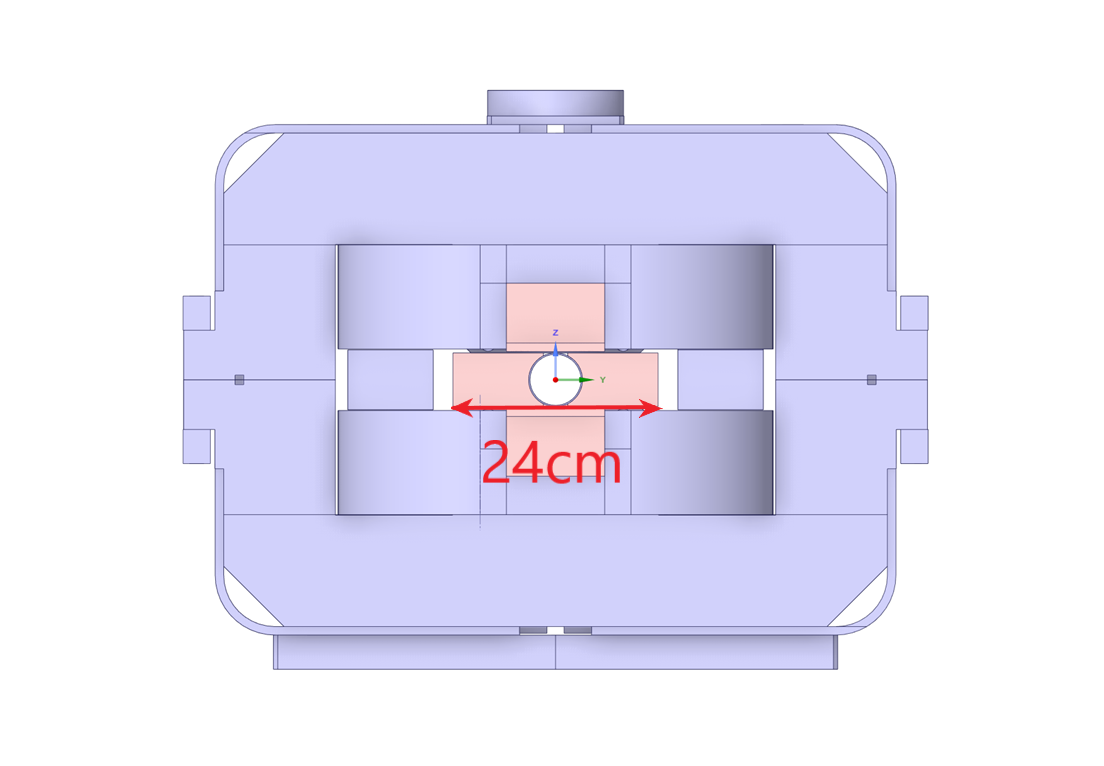}
    \includegraphics*[trim=1.0cm 0.5cm 1.0cm 1.0cm, clip=true,width=0.45\columnwidth]{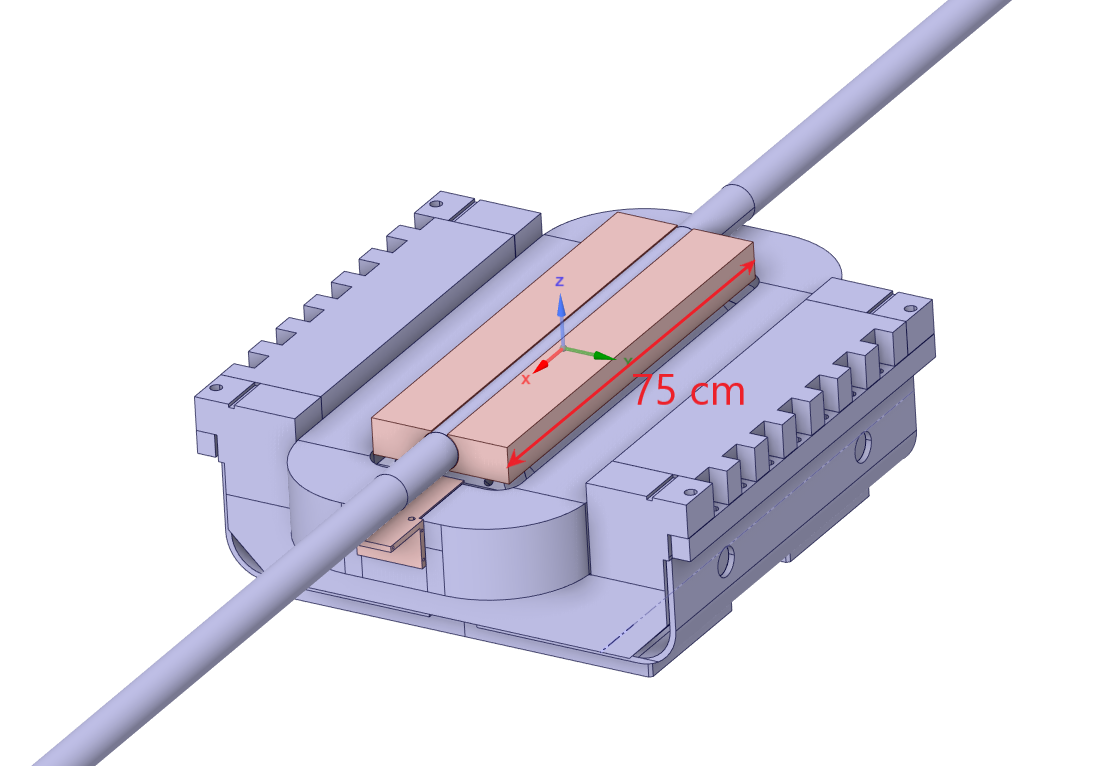}
    \caption{Front (left) and inside (right) view of MBXWS-TAS. The pink blocks are made of tungsten. The shielding to protect its own front coils on the vertical plane is included.}
    \label{fig:TAS}
\end{figure}

\begin{figure}[!h]
    \centering   
    \includegraphics*[trim=1.5cm 0.0cm 1.5cm 0.0cm, clip=true,width=0.95\columnwidth]{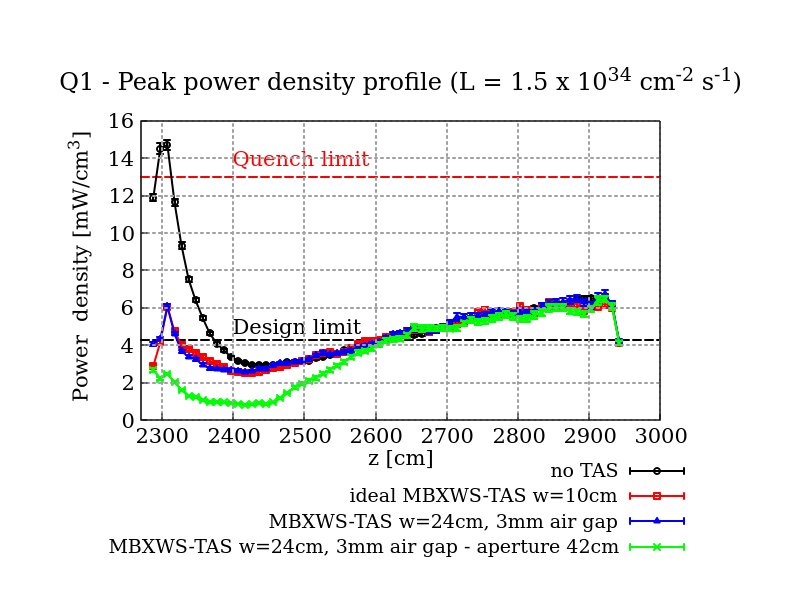}
    \caption{Peak power density along the superconducting coils of the Q1 on the right side with respect to IP8 (at z=0). External vertical crossing and downward polarity of the LHCb spectrometer have been assumed. The black points correspond to the ones reported in green in Fig.~\ref{fig:power_Q1-3_D1}. The red squares represent the scenario featuring an ideal MBXWS-TAS equipped of tungsten in contact with the beam pipe, over a transverse width of 10 cm. The blue triangles correspond to a more realistic MBXWS-TAS design in which a 3~mm gap is left between the pipe and the absorber, to accommodate the bakeout tape, and the absorber transverse width is increased to 24 cm. The green crosses refer to a 10 mm reduction of the vacuum aperture diameter, enabling the shielding extension at smaller radii. Resolution and normalization are as in Fig.~\ref{fig:power_Q1-3_D1}.}
    \label{fig:power_TAS_Q1}
\end{figure}

Building on the experiences of ATLAS and CMS, the most logical choice would be to install a TAS absorber, but the IR8 layout does not allow this due to the presence of the compensator dipole and the detector itself on the right side. 
Considering the very limited space availability, the only effective solution is the integration of an absorbing material inside the yoke of the normal-conducting short compensator (MBXWS) around the beam pipe, converting it into a so called MBXWS-TAS, as shown in Fig.~\ref{fig:TAS}. 
The simulation of a MBXWS-TAS embedding a tungsten shielding surrounding the vacuum chamber all along the magnet length provides a reduction of more than a factor two, well below the quench limit as shown in Fig.~\ref{fig:power_TAS_Q1}. Even if resulting values are still above the design limit, a maximum power density of around $7\,\mathrm{mW/cm^{3}}$ is to be considered as acceptable, as previously pointed out with reference to the Run~2 experience in the ATLAS and CMS insertions. 
Assuming a realistic scenario in which the vacuum pipe is wrapped in a backout tape and therefore the tungsten blocks cannot be in contact with it, no significant loss of shielding effectiveness is observed (see the blue curve in Fig.~\ref{fig:power_TAS_Q1}). Another geometry model has been simulated, reducing the MBXWS-TAS vacuum chamber diameter by 10 mm and increasing the tungsten volume accordingly (see the green curve). This further improves the picture, especially with regard to the total heat load on the cryogenic system, as discussed later.. 
In Fig.~\ref{fig:dose_Q1-3_D1}, the black curve indicates that the MBXWS replacement by the MBXWS-TAS during LS4 solves also the lifetime problem, bringing the Q1 dose well below the respective limit.

\subsubsection{Inner shielding for separation dipole (D1) \label{D1}}
\begin{figure}[!h]
    \centering   
    \includegraphics*[trim=1.5cm 2.5cm 2.5cm 4.0cm, clip=true,width=0.95\columnwidth]{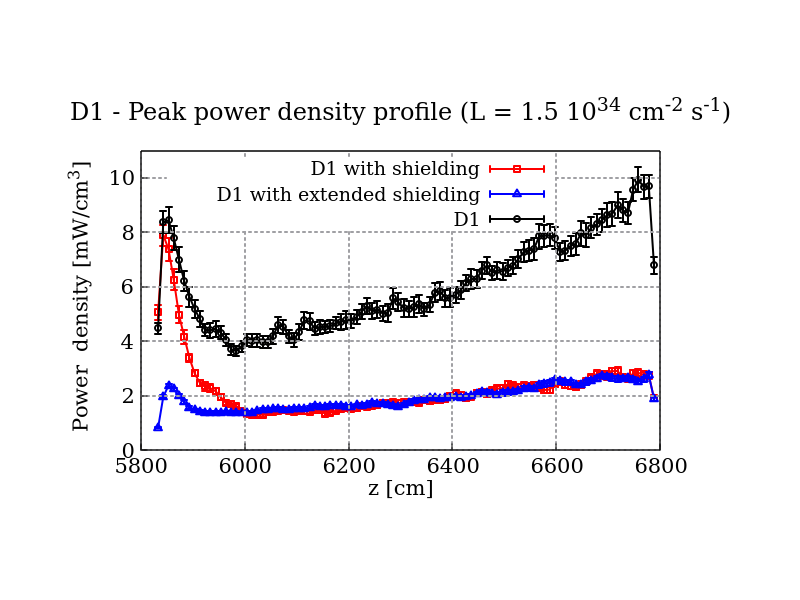}
    \caption{Peak power density along the superconducting coils of the D1 on the right side with respect to IP8 (at z=0). External vertical crossing and downward polarity of the LHCb spectrometer have been assumed. The black points correspond to the ones reported in green in Fig.~\ref{fig:power_Q1-3_D1}. The red squares represent the scenario featuring the proposed replacement of the D1 beam screen allowing for the increase of the stainless steel cold bore wall thickness by {5.5 mm}. The blue triangles are obtained extending upstream the thick cold bore up to the Q3 non-IP end. Resolution and normalization are as in Fig.~\ref{fig:power_Q1-3_D1}.}
    \label{fig:power_D1}
\end{figure}

 As shown in Figs.~\ref{fig:power_Q1-3_D1} and~\ref{fig:dose_Q1-3_D1}, considerable values of deposited energy density would be reached in D1, calling for a mitigation measure such as the design of an internal shielding along the dipole. In fact, the superconducting D1 has a 80 mm coil diameter, larger than the 70 mm one of the triplet quadrupoles, and hosts a wider beam screen. The latter could be replaced by the Q2--Q3 beam screen, if the corresponding mechanical aperture reduction is confirmed not to compromise future injection optics scenarios, offering the margin for increasing the thickness of the enclosing stainless steel cold bore wall. This would serve as a shielding, as it is already the case in Q1, which hosts an even narrower beam screen. The result is that the peak power density at the non-IP-side of D1 is reduced by a factor four, as shown in Fig.~\ref{fig:power_D1}. This constitutes an effective mitigation measure for the non-IP-side peak, also in terms of luminosity lifetime reach (see Fig.~\ref{fig:dose_Q1-3_D1}).
However, the cure for the IP-side power density peak, which is present for both spectrometer configurations, would imply the extension of the thick cold bore along a large fraction of the Q3--D1 interconnect length (see the blue curve of Fig.~\ref{fig:power_D1}), whose implementation has to be further studied. 

\subsubsection{Corrector magnet lifetime \label{correctors}}
Figure~\ref{fig:dose_Q1-3_D1} includes the various corrector magnets of the triplet. They are operated at relatively low current and therefore benefit from a considerable quench margin. Nevertheless, the damage limit of their coil insulator (around 7 MGy~\cite{Corrector_limit}) is much lower than for the main quadrupoles and is predicted to be significantly surpassed by the high order correctors. This is also the case for the ATLAS and CMS insertions, where their possible loss during Run 3 is judged tolerable. 
On the other hand, the orbit corrector dipoles and the skew quadrupole, whose functionality is more critical, are expected to barely reach the limit at the end of the desired luminosity production.

\subsubsection{Total heat load}
The budget of the heat load to be evacuated constitutes one of the bottlenecks in the achievement of the Upgrade II of LHCb.
Table~\ref{tot_load} summarises the power absorbed by some key elements due to the collision debris.
With the current layout, the total power of the right (left) Q1--D1 string is expected to reach $390\,\mathrm{W}$ ($331\,\mathrm{W}$). During Run~3, at the target instantaneous luminosity of the LHCb Upgrade I (7.5 times lower than in the scenario here considered), the load amounts to about 50 W and can be safely removed by the cooling system that has presently a capacity of 140 W. Our study assessed also the effects of the mitigation measures previously discussed to decrease the local energy deposition in the superconducting coils. The load of the Q1 on the right side, which is higher than for the Q1 on the left side, can be decreased from 220 W down to 100 W thanks to a MBXWS-TAS embedding tungsten blocks over a transverse width of 24 cm (see Fig.~\ref{fig:TAS}), provided that the vacuum chamber diameter is reduced to 42 mm (as for the green curve in Fig.~\ref{fig:power_TAS_Q1}). 
In this case, the total power for the whole Q1--D1 string would decrease to 272 W (264 W for the left side) and the new MBXWS-TAS would absorb itself more than twice the heat load of the current MBXWS, namely up to 100 W. Thermo-mechanical calculations indicated that a close-by ventilation system would not be enough to properly lower the temperature of the short compensator vacuum chamber, therefore the outgassing excess would have be handled with additional vacuum pumps.\cite{2896320}.

On the other hand, the aforementioned inner shielding of the separation dipole would collect an additional sizeable amount of power at cryogenic temperature, raising the D1 heat load from 56 W to 96 W and bringing the total heat load of the Q1--D1 string to slightly exceed 300 W, in the presence of the most protective MBXWS-TAS. A possible solution to cope with such an increased load is to replace the current service module with another equivalent to those installed around CMS, assuring a power extraction capacity of 320 W. 


\begin{table*}[!hbt]
   \centering
   \caption{Total heat loads for an instantaneous luminosity of $1.5\cdot10^{34}\,\mathrm{cm}^{-2}\,\mathrm{s}^{-1}$, as a function of the mitigation measures discussed in the text.}
      \begin{ruledtabular}
      \begin{threeparttable}
   \begin{tabular}{lccc}

        \textbf{Magnet} & \multicolumn{3}{c}{\textbf{Total power [W]} }\\
\colrule
        & \textbf{no measures} & \textbf{with MBXWS-TAS} & \textbf{with D1 shielding in addition}\\
\colrule
           \textbf{Right Q1} & 220 & 138/100\tnote{\textbf{a}} & 138/100\tnote{\textbf{a}} \\ 
          \textbf{D1 (down polarity)} & 56 & 56 & 96\\
\colrule
          \textbf{Right triplet and D1 (down polarity)} & 390 & 310$\,/$272\tnote{\textbf{a}}& 350/312\tnote{\textbf{a}}\\ 
          \textbf{Left triplet and D1 (down polarity)} & 331 & 301 $\,/$264\tnote{\textbf{a}}&341/304\tnote{\textbf{a}}\\ 
\colrule
        \textbf{D2 (down polarity)} & 50 & 50 & 50 \\

   \end{tabular}
   \begin{tablenotes}
   \item[\textbf{a}] Assuming a reduced vacuum chamber aperture of $42\,\mathrm{mm}$ diameter in the MBXWS-TAS.
  \end{tablenotes}
   \label{tot_load}
   \end{threeparttable}
      \end{ruledtabular}
\end{table*}

\subsection{Recombination dipole - D2}
\begin{figure}[!tbh]
    \centering
       \begin{overpic}[trim=1.5cm 2.5cm 0.5cm 0.0cm, clip=true,width=0.95\columnwidth]{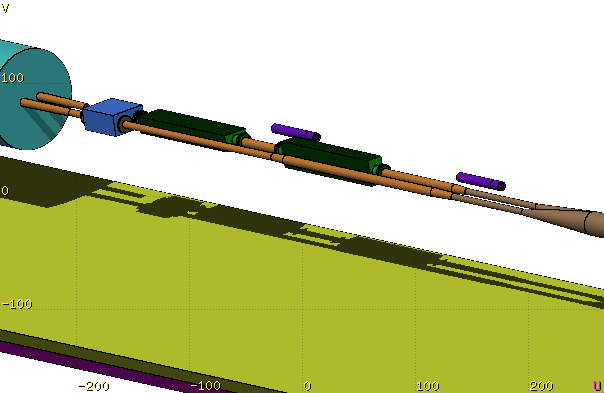}
        \put(130,540){\linethickness{0.3mm}\color{blue}\vector(-0.5,-0.7){70}}
        \put (120,550) {\textbf{\color{blue} D2}}
        \put (70,300) {\textbf{\color{red} TANB}}
        \put (250,420) {\textbf{\color{red} TCTPH}}
        \put (500,360) {\textbf{\color{red} TCTPV}}
        \end{overpic}
    \caption{3D view (from the inside of the ring) of the FLUKA geometry on the left side of IP8 including the recombination dipole D2, the TANB absorber (installed during LS2), and the horizontal and vertical tertiary collimators, which are installed on the incoming (external) beam vacuum chamber.}
    \label{fig:geo_TANB}
\end{figure}

\begin{figure}[!tbh]
    \centering
    \includegraphics[trim=1.5cm 2.5cm 2.5cm 1.8cm, clip=true,width=0.95\columnwidth]{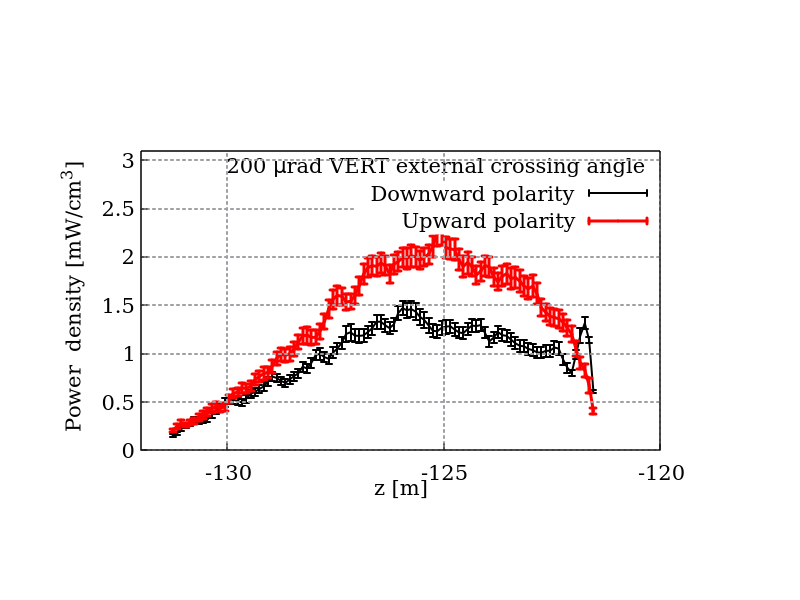}
    \caption{Longitudinal profile of peak power density in the D2 superconducting coils on the left side of IP8 (at z=0). Values are averaged over the cable radial thickness and normalized to $1.5\cdot10^{34}\,\mathrm{cm}^{-2}\,\mathrm{s}^{-1}$. The azimuthal resolution is of $2^{\circ}$. External vertical crossing has been considered for $\sqrt{s}=14\,\mathrm{TeV}$ with either upward (red points) and downward (black points) polarity of the LHCb spectrometer.}
    \label{fig:peak_power_D2}
\end{figure}

The neutral debris with a highly forward angular distribution, mostly neutrons and photons, travels beyond the separation dipole, after which the common vacuum chamber splits into two distinct beam pipes. To protect the subsequent recombination dipole (D2) from these particles, in view of the Upgrade I luminosity increase to $2\cdot10^{33}\,\mathrm{cm}^{-2}\,\mathrm{s}^{-1}$, during LS2 the TANB absorber, covering the empty space between the two pipes, was installed, as illustrated in Fig.~\ref{fig:geo_TANB}.
\begin{figure}[!tbh]
    \centering
    \includegraphics*[trim=3cm 2.5cm 2.5cm 1.8cm, clip=true,width=0.95\columnwidth]{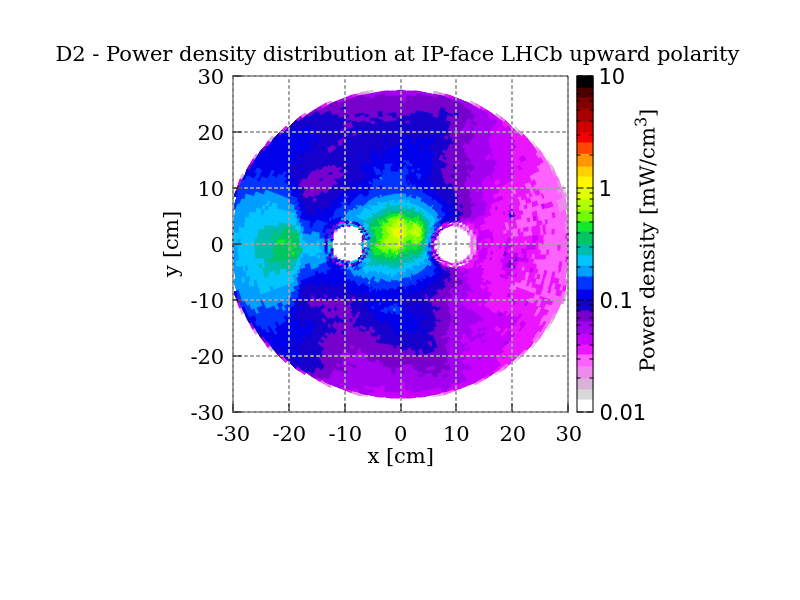}
    \includegraphics*[trim=3cm 2.5cm 2.5cm 1.8cm, clip=true,width=0.95\columnwidth]{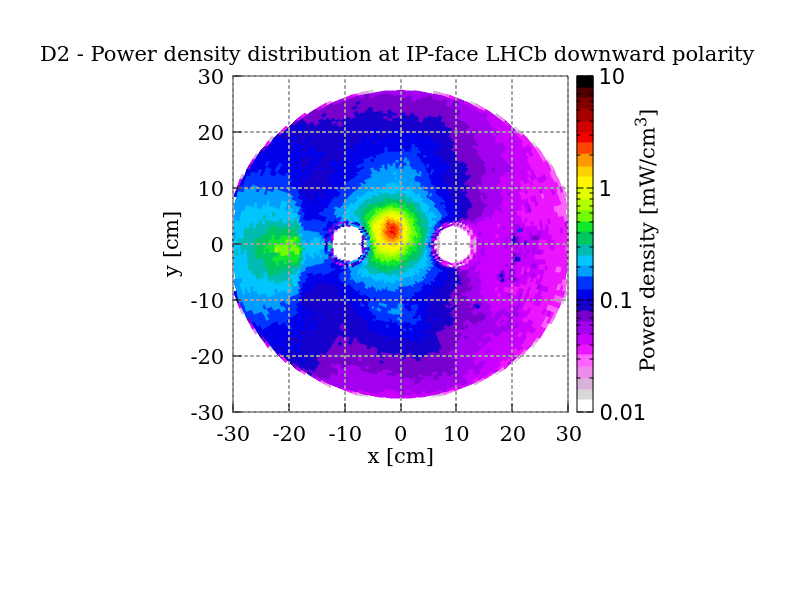}
    \caption{Power density distribution on the IP-face of D2 for upward (top) and downward (bottom) polarity of the LHCb spectrometer in the case of external vertical crossing.}
    \label{fig:power_D2_IPface}
\end{figure}
It turns out to effectively protect the D2 even in the Upgrade II scenario. In fact, Fig.~\ref{fig:peak_power_D2} shows that the peak power density barely exceeds $2\,\mathrm{mW/cm^{3}}$, far below the expected quench limit. 

\begin{figure}[!tbh]
    \centering
    \includegraphics[trim=3cm 2.5cm 2.5cm 1.8cm, clip=true,width=0.95\columnwidth]{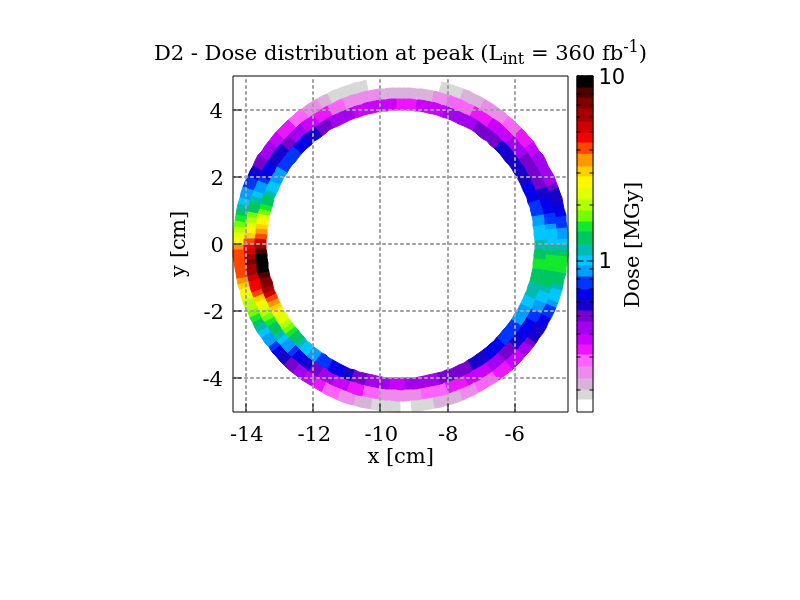}
    \caption{Transverse dose distribution at the longitudinal maximum in the D2 superconducting coils of the outgoing beam on the left side of IP8 for external vertical crossing. An integrated luminosity of $360\,\mathrm{fb^{-1}}$ is assumed to be collected half with either LHCb polarity.}
    \label{fig:dose_D2}
\end{figure}

Fig.~\ref{fig:power_D2_IPface} displays two hot spots on the front face of the D2: a central one between the beam pipes, which is due to the neutral particles from IP8, and one on the inner side, mostly due to positively charged particles. 
The resulting peak in the superconducting coil is localized in the horizontal plane around the outgoing beam pipe, which is the internal one (towards the centre of the LHC ring at negative $x$-values). This is apparent in Fig.~\ref{fig:dose_D2}, indicating that the dose accumulated after $360\,\mathrm{fb^{-1}}$ with external vertical crossing is about 10 MGy. Considering that the additional contribution from the luminosity already collected (mainly with external horizontal crossing) is less than 1 MGy, no radiation induced issue is expected with regard to magnet lifetime. 

Nevertheless, the D2 total load amounts to 50~W, which is higher than the respective HL-LHC value in the ATLAS and CMS insertions. This may require an upgrade of the cryogenic module, which has to evacuate in addition the downstream quadrupole (Q4) load of 5~W. 

Contrary to the present situation, the TANB absorber will require a cooling system to dissipate the absorbed power, which has been evaluated to be {150 W} in the more severe case of downward polarity of the LHCb spectrometer. The currently preferred option envisages cooling plates underneath its bakeout jacket in order to decrease the temperature of the vacuum chambers and to fulfill at the same time the outgassing requirements~\cite{2729331}.

\subsection{Matching section and dispersion suppressor}
Beyond the D2, calculations carried out for the IR8 matching section indicate that the power density and the dose in the superconducting coils of the Q5, Q6 and Q7 quadrupoles remain below $2\,\mathrm{mW/cm^{3}}$ and 3 MGy, respectively. These findings suggest that physics debris collimators (TCL) are not necessary to protect the matching section magnets. 

\begin{figure}[!tbh]
    \centering
    \includegraphics[trim=1cm 1cm 0cm 0cm, clip=true,width=0.95\columnwidth]{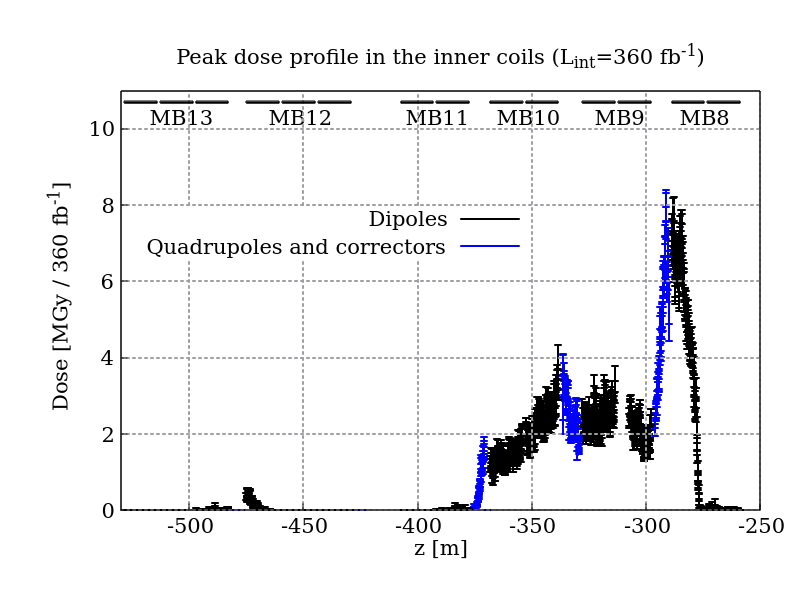}
    \caption{Longitudinal profile of peak dose in the superconducting coils of the magnets of the left DS, assuming an integrated luminosity of $360\,\mathrm{fb}^{-1}$ collected half with either polarity of the LHCb spectrometer, for external vertical crossing.}
    \label{fig:DS_dose}
\end{figure}

Further on, after the Long Straight Section (LSS), particle losses in the dispersion suppressor (DS) are due to diffractive protons coming from the IP with an energy slightly lower than the beam one. They impact the first half-cells (8--10) and then the even half-cells, corresponding to the peaks of the single turn optical dispersion function. The largest power deposition is in the half-cell 8, as shown in Fig.~\ref{fig:DS_dose}. 
The resulting levels are compatible with the operational (quench) and lifetime (dose) limits of the main dipoles and quadrupoles, but raise some concern for the MCBC corrector dipole in the half-cell 8, whose insulator is less resistant (see also Sec.~\ref{correctors}). The predicted dose of 8 MGy suggests to consider the installation of one TCL collimator towards the LSS end, such as to protect the first DS cell.

\section{Conclusions and Outlook}

The Upgrade II of the LHCb experiment, proposed to be implemented by the LHC Run 5 start in 2035, implies the design and installation of protective elements in IR8 to enable the operation at an instantaneous luminosity of $1.5\cdot\,10^{34}\,\mathrm{cm^{−2}\,s^{−1}}$ and the accumulation of an integrated luminosity of about $400\,\mathrm{fb^{-1}}$ by the end of the HL-LHC era. 

In this paper, we have presented extensive modelling of the LHCb upgraded detector and the surrounding LHC machine to explore the impact of this higher luminosity on key elements. The increased luminosity has implications on the risk of magnet quench, the higher cryogenic load, the magnet lifetime, and the radiation to electronics and we make the following conclusions.

\begin{itemize}
\item 
Technological choices for Upgrade II of the LHCb detector described in the Upgrade II TDR \cite{LHCbCollaboration:2776420} are under active investigation within the LHCb collaboration. The basic detector layout will remain similar to that of Run~3, but most of the existing components will have to be replaced to ensure equal or better performance of the detector at 7.5 times higher pile-up by increasing granularity and timing resolution.
Innovative technological solutions for sub-detectors will be required to meet expectations of performance and operational reliability. Judging from the preliminary values for radiation level estimators, the overall increase in luminosity will be a serious concern in terms of radiation damage for certain sub-detectors and is addressed in the R\&D program of new equipment. Performance-driven changes of sub-detector designs will also influence the radiation environment. For example, the introduction of tungsten shielding in the innermost ECAL modules will modify the characteristics of the particle backsplash, inducing effects on adjacent sub-detectors, such as SciFi, Mighty tracker and RICH2.

\item 
Radiation levels in the vicinity of the LHCb experiment are expected to increase by a factor of 5 with respect to the Run~3-4 values, implying the need to protect the electronics to keep SEE rates below an acceptable threshold. One solution we propose that can meet this requirement is a large wall in the experimental cavern.
With our proposed wall design, the HEH fluence in the UX85 can be reduced by a factor of 5, down to  the acceptable values foreseen for Run~3-4. On the other hand, in the US85, the reduction is closer to a factor 4, yielding radiation levels $20\%$ higher than Run~3-4 ones. These radiation levels are manageable. Other less invasive solutions, such as electronics relocation outside UX85 and localized shielding in US85, are currently under investigation.

\item 
The preservation of the magnet functionality of the MBXWS front coils can be assured by a tungsten shielding length of 7 cm.

\item 
The installation of a TAS-like absorber, as the proposed TAS-MBXWS, is required to decrease the power density and dose in the Q1 coils to acceptable levels, as well as to significantly reduce its heat load, such as to be compatible with the upgraded cryogenic system capacity. 

\item 
In the LSS regions, an inner shield is required to reduce the energy deposition peaks in the separation dipole. In particular, its extension along the Q3--D1 interconnection appears to be necessary to cure the peak on the D1 IP face. 

\item 
The installed TANB effectively protects the recombination dipole also in the Upgrade II scenario.
The D2 heat load will rise to 50 W. 

\item 
The TCL collimators do not appear to be necessary for the matching section magnet protection in IR8, except for one collimator at the end of the LSS that would considerably reduce the leakage to the DS and in particular would protect MCBC corrector in the half cell 8 from the insulator degradation risk.
\end{itemize}

In conclusion, we have shown the proposed luminosity could be achievable from a machine protection and radiation level perspective, opening the door to expand the exciting heavy flavour physics program at the HL-LHC. 

\begin{acknowledgments}
This research was supported by the HL-LHC project and the HL-LHC-UK project funded by STFC. This work has been partially sponsored by the Wolfgang Gentner Programme of the German Federal Ministry of Education and Research (grant no. 13E18CHA). We wish to acknowledge the collaboration with the CERN colleagues Riccardo De Maria and Stephane Fartoukh, who provided us with valuable optics input, Francisco Sanchez Galan, who offered fruitful suggestions, and Gerard Ferlin and Johan Bremer for their helpful feedback and support on cryogenics-related aspects.
\end{acknowledgments}


\bibliography{apssamp}

\providecommand{\noopsort}[1]{}\providecommand{\singleletter}[1]{#1}
\begin{thebibliography}{30}%
\makeatletter
\providecommand \@ifxundefined [1]{%
 \@ifx{#1\undefined}
}%
\providecommand \@ifnum [1]{%
 \ifnum #1\expandafter \@firstoftwo
 \else \expandafter \@secondoftwo
 \fi
}%
\providecommand \@ifx [1]{%
 \ifx #1\expandafter \@firstoftwo
 \else \expandafter \@secondoftwo
 \fi
}%
\providecommand \natexlab [1]{#1}%
\providecommand \enquote  [1]{``#1''}%
\providecommand \bibnamefont  [1]{#1}%
\providecommand \bibfnamefont [1]{#1}%
\providecommand \citenamefont [1]{#1}%
\providecommand \href@noop [0]{\@secondoftwo}%
\providecommand \href [0]{\begingroup \@sanitize@url \@href}%
\providecommand \@href[1]{\@@startlink{#1}\@@href}%
\providecommand \@@href[1]{\endgroup#1\@@endlink}%
\providecommand \@sanitize@url [0]{\catcode `\\12\catcode `\$12\catcode
  `\&12\catcode `\#12\catcode `\^12\catcode `\_12\catcode `\%12\relax}%
\providecommand \@@startlink[1]{}%
\providecommand \@@endlink[0]{}%
\providecommand \url  [0]{\begingroup\@sanitize@url \@url }%
\providecommand \@url [1]{\endgroup\@href {#1}{\urlprefix }}%
\providecommand \urlprefix  [0]{URL }%
\providecommand \Eprint [0]{\href }%
\providecommand \doibase [0]{http://dx.doi.org/}%
\providecommand \selectlanguage [0]{\@gobble}%
\providecommand \bibinfo  [0]{\@secondoftwo}%
\providecommand \bibfield  [0]{\@secondoftwo}%
\providecommand \translation [1]{[#1]}%
\providecommand \BibitemOpen [0]{}%
\providecommand \bibitemStop [0]{}%
\providecommand \bibitemNoStop [0]{.\EOS\space}%
\providecommand \EOS [0]{\spacefactor3000\relax}%
\providecommand \BibitemShut  [1]{\csname bibitem#1\endcsname}%
\let\auto@bib@innerbib\@empty
\bibitem [{\citenamefont {Collaboration}(2008)}]{Alves:1129809}%
  \BibitemOpen
  \bibfield  {author} {\bibinfo {author} {\bibfnamefont {L.}~\bibnamefont
  {Collaboration}},\ }\href {\doibase 10.1088/1748-0221/3/08/S08005} {\bibfield
   {journal} {\bibinfo  {journal} {JINST}\ }\textbf {\bibinfo {volume} {3}},\
  \bibinfo {pages} {S08005} (\bibinfo {year} {2008})},\ \bibinfo {note} {also
  published by CERN Geneva in 2010}\BibitemShut {NoStop}%
\bibitem [{\citenamefont {CERN}(2022)}]{CERN_LHC_schedule}%
  \BibitemOpen
  \bibfield  {author} {\bibinfo {author} {\bibnamefont {CERN}},\ }\href@noop {}
  {\enquote {\bibinfo {title} {{Longer term LHC schedule}},}\ }\bibinfo
  {howpublished}
  {\url{http://lhc-commissioning.web.cern.ch/schedule/LHC-long-term.htm}}
  (\bibinfo {year} {accessed: 2022})\BibitemShut {NoStop}%
\bibitem [{\citenamefont {{LHCb
  Collaboration}}(2021)}]{LHCbCollaboration:2776420}%
  \BibitemOpen
  \bibfield  {author} {\bibinfo {author} {\bibnamefont {{LHCb
  Collaboration}}},\ }\href {https://cds.cern.ch/record/2776420} {\emph
  {\bibinfo {title} {{Framework TDR for the LHCb Upgrade II - Opportunities in
  flavour physics, and beyond, in the HL-LHC era}}}},\ \bibinfo {type} {Tech.
  Rep.}\ \bibinfo {number} {CERN-LHCC-2021-012, LHCB-TDR-023}\ (\bibinfo
  {institution} {CERN},\ \bibinfo {address} {Geneva},\ \bibinfo {year}
  {2021})\BibitemShut {NoStop}%
\bibitem [{\citenamefont {Efthymiopoulos}\ \emph {et~al.}(2018)\citenamefont
  {Efthymiopoulos}, \citenamefont {Arduini}, \citenamefont {Baglin},
  \citenamefont {Burkhardt}, \citenamefont {Cerutti}, \citenamefont {Claudet}
  \emph {et~al.}}]{Efthymiopoulos:2319258}%
  \BibitemOpen
  \bibfield  {author} {\bibinfo {author} {\bibfnamefont {I.}~\bibnamefont
  {Efthymiopoulos}}, \bibinfo {author} {\bibfnamefont {G.}~\bibnamefont
  {Arduini}}, \bibinfo {author} {\bibfnamefont {V.}~\bibnamefont {Baglin}},
  \bibinfo {author} {\bibfnamefont {H.}~\bibnamefont {Burkhardt}}, \bibinfo
  {author} {\bibfnamefont {F.}~\bibnamefont {Cerutti}}, \bibinfo {author}
  {\bibfnamefont {S.}~\bibnamefont {Claudet}},  \emph {et~al.},\ }\href
  {https://cds.cern.ch/record/2319258} {\emph {\bibinfo {title} {{LHCb Upgrades
  and operation at $10^{34} cm^{-2} s^{-1}$ luminosity – A first study}}}},\
  \bibinfo {type} {Tech. Rep.}\ \bibinfo {number} {{CERN-ACC-NOTE-2018-0038}}\
  (\bibinfo  {institution} {CERN},\ \bibinfo {year} {2018})\BibitemShut
  {NoStop}%
\bibitem [{\citenamefont {Ciccotelli}\ \emph {et~al.}(2023)\citenamefont
  {Ciccotelli}, \citenamefont {Appleby}, \citenamefont {Cerutti}, \citenamefont
  {Bilko}, \citenamefont {Esposito}, \citenamefont {Alia}, \citenamefont
  {Lechner},\ and\ \citenamefont {Tsinganis}}]{PhysRevAccelBeams.26.061002}%
  \BibitemOpen
  \bibfield  {author} {\bibinfo {author} {\bibfnamefont {A.}~\bibnamefont
  {Ciccotelli}}, \bibinfo {author} {\bibfnamefont {R.~B.}\ \bibnamefont
  {Appleby}}, \bibinfo {author} {\bibfnamefont {F.}~\bibnamefont {Cerutti}},
  \bibinfo {author} {\bibfnamefont {K.}~\bibnamefont {Bilko}}, \bibinfo
  {author} {\bibfnamefont {L.~S.}\ \bibnamefont {Esposito}}, \bibinfo {author}
  {\bibfnamefont {R.~G.}\ \bibnamefont {Alia}}, \bibinfo {author}
  {\bibfnamefont {A.}~\bibnamefont {Lechner}}, \ and\ \bibinfo {author}
  {\bibfnamefont {A.}~\bibnamefont {Tsinganis}},\ }\href {\doibase
  10.1103/PhysRevAccelBeams.26.061002} {\bibfield  {journal} {\bibinfo
  {journal} {Phys. Rev. Accel. Beams}\ }\textbf {\bibinfo {volume} {26}},\
  \bibinfo {pages} {061002} (\bibinfo {year} {2023})}\BibitemShut {NoStop}%
\bibitem [{\citenamefont {Albrecht}\ \emph {et~al.}(2019)\citenamefont
  {Albrecht}, \citenamefont {Charles}, \citenamefont {Dufour}, \citenamefont
  {Needham}, \citenamefont {Parkes}, \citenamefont {Passaleva} \emph
  {et~al.}}]{Albrecht:2653011}%
  \BibitemOpen
  \bibfield  {author} {\bibinfo {author} {\bibfnamefont {J.}~\bibnamefont
  {Albrecht}}, \bibinfo {author} {\bibfnamefont {M.~J.}\ \bibnamefont
  {Charles}}, \bibinfo {author} {\bibfnamefont {L.}~\bibnamefont {Dufour}},
  \bibinfo {author} {\bibfnamefont {M.~D.}\ \bibnamefont {Needham}}, \bibinfo
  {author} {\bibfnamefont {C.}~\bibnamefont {Parkes}}, \bibinfo {author}
  {\bibfnamefont {G.}~\bibnamefont {Passaleva}},  \emph {et~al.},\ }\href
  {https://cds.cern.ch/record/2653011} {\emph {\bibinfo {title} {{Luminosity
  scenarios for LHCb Upgrade II}}}},\ \bibinfo {type} {Tech. Rep.}\ \bibinfo
  {number} {LHCb-PUB-2019-001}\ (\bibinfo  {institution} {CERN},\ \bibinfo
  {address} {Geneva},\ \bibinfo {year} {2019})\BibitemShut {NoStop}%
\bibitem [{\citenamefont {Tom\'as}(2023)}]{triplet}%
  \BibitemOpen
  \bibfield  {author} {\bibinfo {author} {\bibfnamefont {R.}~\bibnamefont
  {Tom\'as}},\ }\href@noop {} {\enquote {\bibinfo {title} {{{Private
  communication}}},}\ } (\bibinfo {year} {2023})\BibitemShut {NoStop}%
\bibitem [{\citenamefont {CERN}()}]{FLUKA}%
  \BibitemOpen
  \bibfield  {author} {\bibinfo {author} {\bibnamefont {CERN}},\ }\href@noop {}
  {\enquote {\bibinfo {title} {Fluka website},}\ }\bibinfo {howpublished}
  {\url{https://fluka.cern}}\BibitemShut {NoStop}%
\bibitem [{\citenamefont {Ahdida}\ \emph {et~al.}(2022)\citenamefont {Ahdida},
  \citenamefont {Bozzato}, \citenamefont {Calzolari}, \citenamefont {Cerutti},
  \citenamefont {Charitonidis}, \citenamefont {Cimmino} \emph
  {et~al.}}]{FLUKA2021}%
  \BibitemOpen
  \bibfield  {author} {\bibinfo {author} {\bibfnamefont {C.}~\bibnamefont
  {Ahdida}}, \bibinfo {author} {\bibfnamefont {D.}~\bibnamefont {Bozzato}},
  \bibinfo {author} {\bibfnamefont {D.}~\bibnamefont {Calzolari}}, \bibinfo
  {author} {\bibfnamefont {F.}~\bibnamefont {Cerutti}}, \bibinfo {author}
  {\bibfnamefont {N.}~\bibnamefont {Charitonidis}}, \bibinfo {author}
  {\bibfnamefont {A.}~\bibnamefont {Cimmino}},  \emph {et~al.},\ }\href
  {\doibase 10.3389/fphy.2021.788253} {\bibfield  {journal} {\bibinfo
  {journal} {Frontiers in Physics}\ }\textbf {\bibinfo {volume} {9}} (\bibinfo
  {year} {2022}),\ 10.3389/fphy.2021.788253}\BibitemShut {NoStop}%
\bibitem [{\citenamefont {Battistoni}\ \emph {et~al.}(2015)\citenamefont
  {Battistoni}, \citenamefont {Boehlen}, \citenamefont {Cerutti}, \citenamefont
  {Chin}, \citenamefont {Esposito}, \citenamefont {Fass{\`o}} \emph
  {et~al.}}]{FLUKA:2015}%
  \BibitemOpen
  \bibfield  {author} {\bibinfo {author} {\bibfnamefont {G.}~\bibnamefont
  {Battistoni}}, \bibinfo {author} {\bibfnamefont {T.}~\bibnamefont {Boehlen}},
  \bibinfo {author} {\bibfnamefont {F.}~\bibnamefont {Cerutti}}, \bibinfo
  {author} {\bibfnamefont {P.~W.}\ \bibnamefont {Chin}}, \bibinfo {author}
  {\bibfnamefont {L.~S.}\ \bibnamefont {Esposito}}, \bibinfo {author}
  {\bibfnamefont {A.}~\bibnamefont {Fass{\`o}}},  \emph {et~al.},\ }\href
  {\doibase 10.1016/j.anucene.2014.11.007} {\bibfield  {journal} {\bibinfo
  {journal} {Ann. Nucl. Energy}\ }\textbf {\bibinfo {volume} {82}},\ \bibinfo
  {pages} {10} (\bibinfo {year} {2015})}\BibitemShut {NoStop}%
\bibitem [{\citenamefont {Mereghetti}\ \emph {et~al.}(2012)\citenamefont
  {Mereghetti}, \citenamefont {Boccone}, \citenamefont {Cerutti}, \citenamefont
  {Versaci},\ and\ \citenamefont {Vlachoudis}}]{Mereghetti:2012zz}%
  \BibitemOpen
  \bibfield  {author} {\bibinfo {author} {\bibfnamefont {A.}~\bibnamefont
  {Mereghetti}}, \bibinfo {author} {\bibfnamefont {V.}~\bibnamefont {Boccone}},
  \bibinfo {author} {\bibfnamefont {F.}~\bibnamefont {Cerutti}}, \bibinfo
  {author} {\bibfnamefont {R.}~\bibnamefont {Versaci}}, \ and\ \bibinfo
  {author} {\bibfnamefont {V.}~\bibnamefont {Vlachoudis}},\ }\bibfield
  {booktitle} {\emph {\bibinfo {booktitle} {{Proceedings, 3rd International
  Conference on Particle accelerator (IPAC 2012): New Orleans, USA, May 2-25,
  2012}}},\ }\href@noop {} {\bibfield  {journal} {\bibinfo  {journal} {Conf.
  Proc.}\ }\textbf {\bibinfo {volume} {C1205201}},\ \bibinfo {pages} {2687}
  (\bibinfo {year} {2012})}\BibitemShut {NoStop}%
\bibitem [{\citenamefont {Vlachoudis}(2009)}]{Vlachoudis:2749540}%
  \BibitemOpen
  \bibfield  {author} {\bibinfo {author} {\bibfnamefont {V.}~\bibnamefont
  {Vlachoudis}},\ }\href {https://cds.cern.ch/record/2749540} {\bibfield
  {journal} {\bibinfo  {journal} {{International Conference on Mathematics,
  Computational Methods \& Reactor Physics, Saragota Springs, New York, 3-7 May
  2009}}\ ,\ \bibinfo {pages} {790}} (\bibinfo {year} {2009})}\BibitemShut
  {NoStop}%
\bibitem [{\citenamefont {{ATLAS~Collaboration}}(2016)}]{ATLAS:2016ygv}%
  \BibitemOpen
  \bibfield  {author} {\bibinfo {author} {\bibnamefont
  {{ATLAS~Collaboration}}},\ }\href {\doibase 10.1103/PhysRevLett.117.182002}
  {\bibfield  {journal} {\bibinfo  {journal} {Phys. Rev. Lett.}\ }\textbf
  {\bibinfo {volume} {117}},\ \bibinfo {pages} {182002} (\bibinfo {year}
  {2016})},\ \Eprint {http://arxiv.org/abs/1606.02625} {arXiv:1606.02625
  [hep-ex]} \BibitemShut {NoStop}%
\bibitem [{\citenamefont {Brugger}\ \emph {et~al.}(2015)\citenamefont
  {Brugger}, \citenamefont {Cerutti},\ and\ \citenamefont
  {Esposito}}]{Brugger:2131739}%
  \BibitemOpen
  \bibfield  {author} {\bibinfo {author} {\bibfnamefont {M.}~\bibnamefont
  {Brugger}}, \bibinfo {author} {\bibfnamefont {F.}~\bibnamefont {Cerutti}}, \
  and\ \bibinfo {author} {\bibfnamefont {L.~S.}\ \bibnamefont {Esposito}},\
  }\href {\doibase 10.1142/9789814675475_0010} {\bibfield  {journal} {\bibinfo
  {journal} {Adv. Ser. Dir. High Energy Phys.}\ }\textbf {\bibinfo {volume}
  {24}},\ \bibinfo {pages} {177} (\bibinfo {year} {2015})}\BibitemShut
  {NoStop}%
\bibitem [{\citenamefont {Lerner}\ \emph {et~al.}(2020)\citenamefont {Lerner}
  \emph {et~al.}}]{Lerner:2302154}%
  \BibitemOpen
  \bibfield  {author} {\bibinfo {author} {\bibfnamefont {G.}~\bibnamefont
  {Lerner}} \emph {et~al.},\ }\href {https://edms.cern.ch/document/2302154/1.0}
  {\emph {\bibinfo {title} {{HL-LHC Radiation level specification
  document}}}},\ \bibinfo {type} {Tech. Rep.}\ \bibinfo {number} {{2302154
  v.1.0,LHC-N-ES-0001 v.1.0}}\ (\bibinfo  {institution} {CERN},\ \bibinfo
  {year} {2020})\BibitemShut {NoStop}%
\bibitem [{\citenamefont {Calvo~Gomez}\ \emph {et~al.}(2017)\citenamefont
  {Calvo~Gomez}, \citenamefont {Corti}, \citenamefont {Joram}, \citenamefont
  {Karacson}, \citenamefont {Lindner}, \citenamefont {Saputi}, \citenamefont
  {Schindler}, \citenamefont {Thomas},\ and\ \citenamefont
  {Uwer}}]{CalvoGomez:2268537}%
  \BibitemOpen
  \bibfield  {author} {\bibinfo {author} {\bibfnamefont {M.}~\bibnamefont
  {Calvo~Gomez}}, \bibinfo {author} {\bibfnamefont {G.}~\bibnamefont {Corti}},
  \bibinfo {author} {\bibfnamefont {C.}~\bibnamefont {Joram}}, \bibinfo
  {author} {\bibfnamefont {M.}~\bibnamefont {Karacson}}, \bibinfo {author}
  {\bibfnamefont {R.}~\bibnamefont {Lindner}}, \bibinfo {author} {\bibfnamefont
  {A.}~\bibnamefont {Saputi}}, \bibinfo {author} {\bibfnamefont
  {H.}~\bibnamefont {Schindler}}, \bibinfo {author} {\bibfnamefont
  {E.}~\bibnamefont {Thomas}}, \ and\ \bibinfo {author} {\bibfnamefont
  {U.}~\bibnamefont {Uwer}},\ }\href {https://cds.cern.ch/record/2268537}
  {\emph {\bibinfo {title} {{Studies of a neutron shielding for the upgraded
  LHCb detector}}}},\ \bibinfo {type} {Tech. Rep.}\ (\bibinfo  {institution}
  {CERN},\ \bibinfo {address} {Geneva},\ \bibinfo {year} {2017})\BibitemShut
  {NoStop}%
\bibitem [{\citenamefont {Calvo~Gomez}\ \emph {et~al.}(2023)\citenamefont
  {Calvo~Gomez}, \citenamefont {Corti}, \citenamefont {Joram}, \citenamefont
  {Karacson}, \citenamefont {Lindner}, \citenamefont {Saputi}, \citenamefont
  {Schindler}, \citenamefont {Thomas},\ and\ \citenamefont
  {Uwer}}]{CalvoGomez:2869609}%
  \BibitemOpen
  \bibfield  {author} {\bibinfo {author} {\bibfnamefont {M.}~\bibnamefont
  {Calvo~Gomez}}, \bibinfo {author} {\bibfnamefont {G.}~\bibnamefont {Corti}},
  \bibinfo {author} {\bibfnamefont {C.}~\bibnamefont {Joram}}, \bibinfo
  {author} {\bibfnamefont {M.}~\bibnamefont {Karacson}}, \bibinfo {author}
  {\bibfnamefont {R.}~\bibnamefont {Lindner}}, \bibinfo {author} {\bibfnamefont
  {A.}~\bibnamefont {Saputi}}, \bibinfo {author} {\bibfnamefont
  {H.}~\bibnamefont {Schindler}}, \bibinfo {author} {\bibfnamefont
  {E.}~\bibnamefont {Thomas}}, \ and\ \bibinfo {author} {\bibfnamefont
  {U.}~\bibnamefont {Uwer}},\ }\href {https://cds.cern.ch/record/2869609}
  {\emph {\bibinfo {title} {{Studies of a neutron shielding for the upgraded
  LHCb detector}}}},\ \bibinfo {type} {Tech. Rep.}\ (\bibinfo  {institution}
  {CERN},\ \bibinfo {address} {Geneva},\ \bibinfo {year} {2023})\ \bibinfo
  {note} {submitted}\BibitemShut {NoStop}%
\bibitem [{\citenamefont {Karacson}(2016)}]{Karacson:2243499}%
  \BibitemOpen
  \bibfield  {author} {\bibinfo {author} {\bibfnamefont {M.}~\bibnamefont
  {Karacson}},\ }\emph {\bibinfo {title} {{Evaluation of the Radiation
  Environment of the LHCb Experiment}}},\ \href
  {https://cds.cern.ch/record/2243499} {Ph.D. thesis},\ \bibinfo  {school}
  {{Vienna, Tech. U.}} (\bibinfo {year} {2016}),\ \bibinfo {note}
  {{CERN-THESIS-2016-246}}\BibitemShut {NoStop}%
\bibitem [{\citenamefont {Spiezia}\ \emph {et~al.}(2014)\citenamefont
  {Spiezia}, \citenamefont {Peronnard}, \citenamefont {Masi}, \citenamefont
  {Brugger}, \citenamefont {Brucoli}, \citenamefont {Danzeca} \emph
  {et~al.}}]{Spiezia:2011660}%
  \BibitemOpen
  \bibfield  {author} {\bibinfo {author} {\bibfnamefont {G.}~\bibnamefont
  {Spiezia}}, \bibinfo {author} {\bibfnamefont {P.}~\bibnamefont {Peronnard}},
  \bibinfo {author} {\bibfnamefont {A.}~\bibnamefont {Masi}}, \bibinfo {author}
  {\bibfnamefont {M.}~\bibnamefont {Brugger}}, \bibinfo {author} {\bibfnamefont
  {M.}~\bibnamefont {Brucoli}}, \bibinfo {author} {\bibfnamefont
  {S.}~\bibnamefont {Danzeca}},  \emph {et~al.},\ }\href {\doibase
  10.1109/TNS.2014.2365046} {\bibfield  {journal} {\bibinfo  {journal} {IEEE
  Trans. Nucl. Sci.}\ }\textbf {\bibinfo {volume} {61}},\ \bibinfo {pages}
  {3424} (\bibinfo {year} {2014})}\BibitemShut {NoStop}%
\bibitem [{\citenamefont {Aielli}\ and\ \citenamefont
  {others"}(2020)}]{Aielli_2020}%
  \BibitemOpen
  \bibfield  {author} {\bibinfo {author} {\bibfnamefont {G.}~\bibnamefont
  {Aielli}}\ and\ \bibinfo {author} {\bibnamefont {others"}},\ }\href {\doibase
  10.1140/epjc/s10052-020-08711-3} {\bibfield  {journal} {\bibinfo  {journal}
  {The European Physical Journal C}\ }\textbf {\bibinfo {volume} {80}}
  (\bibinfo {year} {2020}),\ 10.1140/epjc/s10052-020-08711-3}\BibitemShut
  {NoStop}%
\bibitem [{\citenamefont {Dey}\ \emph {et~al.}(2019)\citenamefont {Dey},
  \citenamefont {Lee}, \citenamefont {Coco},\ and\ \citenamefont
  {Moon}}]{dey2019background}%
  \BibitemOpen
  \bibfield  {author} {\bibinfo {author} {\bibfnamefont {B.}~\bibnamefont
  {Dey}}, \bibinfo {author} {\bibfnamefont {J.}~\bibnamefont {Lee}}, \bibinfo
  {author} {\bibfnamefont {V.}~\bibnamefont {Coco}}, \ and\ \bibinfo {author}
  {\bibfnamefont {C.-S.}\ \bibnamefont {Moon}},\ }\href@noop {} {\enquote
  {\bibinfo {title} {{Background studies for the CODEX-b experiment:
  measurements and simulation}},}\ } (\bibinfo {year} {2019}),\ \Eprint
  {http://arxiv.org/abs/1912.03846} {arXiv:1912.03846 [physics.ins-det]}
  \BibitemShut {NoStop}%
\bibitem [{\citenamefont {Zimmaro}\ \emph {et~al.}(2022)\citenamefont
  {Zimmaro}, \citenamefont {Ferraro}, \citenamefont {Boch}, \citenamefont
  {Saigne}, \citenamefont {Garcia~Alia}, \citenamefont {Brucoli}, \citenamefont
  {Masi},\ and\ \citenamefont {Danzeca}}]{Zimmaro:2823933}%
  \BibitemOpen
  \bibfield  {author} {\bibinfo {author} {\bibfnamefont {A.}~\bibnamefont
  {Zimmaro}}, \bibinfo {author} {\bibfnamefont {R.}~\bibnamefont {Ferraro}},
  \bibinfo {author} {\bibfnamefont {J.}~\bibnamefont {Boch}}, \bibinfo {author}
  {\bibfnamefont {F.}~\bibnamefont {Saigne}}, \bibinfo {author} {\bibfnamefont
  {R.}~\bibnamefont {Garcia~Alia}}, \bibinfo {author} {\bibfnamefont
  {M.}~\bibnamefont {Brucoli}}, \bibinfo {author} {\bibfnamefont
  {A.}~\bibnamefont {Masi}}, \ and\ \bibinfo {author} {\bibfnamefont
  {S.}~\bibnamefont {Danzeca}},\ }\href {\doibase 10.1109/TNS.2022.3158527}
  {\bibfield  {journal} {\bibinfo  {journal} {IEEE Trans. Nucl. Sci.}\ }\textbf
  {\bibinfo {volume} {69}},\ \bibinfo {pages} {1642} (\bibinfo {year}
  {2022})}\BibitemShut {NoStop}%
\bibitem [{\citenamefont {Poikela}\ \emph {et~al.}(2014)\citenamefont
  {Poikela}, \citenamefont {Plosila}, \citenamefont {Westerlund}, \citenamefont
  {Campbell}, \citenamefont {Gaspari}, \citenamefont {Llopart}, \citenamefont
  {Gromov}, \citenamefont {Kluit}, \citenamefont {Beuzekom}, \citenamefont
  {Zappon}, \citenamefont {Zivkovic}, \citenamefont {Brezina}, \citenamefont
  {Desch}, \citenamefont {Fu},\ and\ \citenamefont {Kruth}}]{Poikela2014}%
  \BibitemOpen
  \bibfield  {author} {\bibinfo {author} {\bibfnamefont {T.}~\bibnamefont
  {Poikela}}, \bibinfo {author} {\bibfnamefont {J.}~\bibnamefont {Plosila}},
  \bibinfo {author} {\bibfnamefont {T.}~\bibnamefont {Westerlund}}, \bibinfo
  {author} {\bibfnamefont {M.}~\bibnamefont {Campbell}}, \bibinfo {author}
  {\bibfnamefont {M.~D.}\ \bibnamefont {Gaspari}}, \bibinfo {author}
  {\bibfnamefont {X.}~\bibnamefont {Llopart}}, \bibinfo {author} {\bibfnamefont
  {V.}~\bibnamefont {Gromov}}, \bibinfo {author} {\bibfnamefont
  {R.}~\bibnamefont {Kluit}}, \bibinfo {author} {\bibfnamefont {M.~v.}\
  \bibnamefont {Beuzekom}}, \bibinfo {author} {\bibfnamefont {F.}~\bibnamefont
  {Zappon}}, \bibinfo {author} {\bibfnamefont {V.}~\bibnamefont {Zivkovic}},
  \bibinfo {author} {\bibfnamefont {C.}~\bibnamefont {Brezina}}, \bibinfo
  {author} {\bibfnamefont {K.}~\bibnamefont {Desch}}, \bibinfo {author}
  {\bibfnamefont {Y.}~\bibnamefont {Fu}}, \ and\ \bibinfo {author}
  {\bibfnamefont {A.}~\bibnamefont {Kruth}},\ }\href {\doibase
  10.1088/1748-0221/9/05/C05013} {\bibfield  {journal} {\bibinfo  {journal}
  {Journal of Instrumentation}\ }\textbf {\bibinfo {volume} {9}},\ \bibinfo
  {pages} {C05013} (\bibinfo {year} {2014})}\BibitemShut {NoStop}%
\bibitem [{\citenamefont {Workman}\ \emph {et~al.}(2022)\citenamefont {Workman}
  \emph {et~al.}}]{Workman:2022ynf}%
  \BibitemOpen
  \bibfield  {author} {\bibinfo {author} {\bibfnamefont {R.~L.}\ \bibnamefont
  {Workman}} \emph {et~al.} (\bibinfo {collaboration} {Particle Data Group}),\
  }\href {\doibase 10.1093/ptep/ptac097} {\bibfield  {journal} {\bibinfo
  {journal} {PTEP}\ }\textbf {\bibinfo {volume} {2022}},\ \bibinfo {pages}
  {083C01} (\bibinfo {year} {2022})}\BibitemShut {NoStop}%
\bibitem [{\citenamefont {Fessia}\ and\ \citenamefont
  {Cerutti}(2019)}]{1952741}%
  \BibitemOpen
  \bibfield  {author} {\bibinfo {author} {\bibfnamefont {P.}~\bibnamefont
  {Fessia}}\ and\ \bibinfo {author} {\bibfnamefont {F.}~\bibnamefont
  {Cerutti}},\ }\href@noop {} {\emph {\bibinfo {title} {{Radiation levels of
  MBW and MQW}}}},\ \bibinfo {type} {Tech. Rep.}\ \bibinfo {number} {1952741
  v.1}\ (\bibinfo  {institution} {CERN},\ \bibinfo {year} {2019})\BibitemShut
  {NoStop}%
\bibitem [{\citenamefont {Mokhov}\ \emph {et~al.}(2003)\citenamefont {Mokhov},
  \citenamefont {Rakhno}, \citenamefont {Kerby},\ and\ \citenamefont
  {Strait}}]{Mokhov:613167}%
  \BibitemOpen
  \bibfield  {author} {\bibinfo {author} {\bibfnamefont {N.~V.}\ \bibnamefont
  {Mokhov}}, \bibinfo {author} {\bibfnamefont {I.~L.}\ \bibnamefont {Rakhno}},
  \bibinfo {author} {\bibfnamefont {J.~S.}\ \bibnamefont {Kerby}}, \ and\
  \bibinfo {author} {\bibfnamefont {J.~B.}\ \bibnamefont {Strait}} (\bibinfo
  {collaboration} {CERN-US LHC Construction Collaboration}),\ }\href
  {https://cds.cern.ch/record/613167} {\emph {\bibinfo {title} {{Protecting LHC
  IP1/IP5 Components Against Radiation Resulting from Colliding Beam
  Interactions}}}},\ \bibinfo {type} {Tech. Rep.}\ \bibinfo {number}
  {FERMILAB-FN-0732, CERN-LHC-Project-Report-633}\ (\bibinfo  {institution}
  {CERN},\ \bibinfo {address} {Geneva},\ \bibinfo {year} {2003})\BibitemShut
  {NoStop}%
\bibitem [{\citenamefont {Tavlet}\ \emph {et~al.}(1998)\citenamefont {Tavlet},
  \citenamefont {Fontaine},\ and\ \citenamefont {Schonbacher}}]{Tavlet}%
  \BibitemOpen
  \bibfield  {author} {\bibinfo {author} {\bibfnamefont {M.}~\bibnamefont
  {Tavlet}}, \bibinfo {author} {\bibfnamefont {A.}~\bibnamefont {Fontaine}}, \
  and\ \bibinfo {author} {\bibfnamefont {H.}~\bibnamefont {Schonbacher}},\
  }\href {https://cds.cern.ch/record/357576/files/CERN-98-01.pdf} {\emph
  {\bibinfo {title} {{Compilation of radiation damage test data}}}},\ \bibinfo
  {type} {Tech. Rep.}\ \bibinfo {number} {98-01}\ (\bibinfo  {institution}
  {CERN},\ \bibinfo {address} {Geneva},\ \bibinfo {year} {1998})\BibitemShut
  {NoStop}%
\bibitem [{\citenamefont {Karppinen}()}]{Corrector_limit}%
  \BibitemOpen
  \bibfield  {author} {\bibinfo {author} {\bibfnamefont {M.}~\bibnamefont
  {Karppinen}},\ }\href@noop {} {}\bibinfo {howpublished} {private
  communication}\BibitemShut {NoStop}%
\bibitem [{\citenamefont {Wehrle}(2023)}]{2896320}%
  \BibitemOpen
  \bibfield  {author} {\bibinfo {author} {\bibfnamefont {M.}~\bibnamefont
  {Wehrle}},\ }\href@noop {} {\emph {\bibinfo {title} {{Thermic FEA of MBXWS as
  TAS absorber - LHCb Upgrade II}}}},\ \bibinfo {type} {Tech. Rep.}\ \bibinfo
  {number} {2896320 v.0.1}\ (\bibinfo  {institution} {CERN},\ \bibinfo {year}
  {2023})\BibitemShut {NoStop}%
\bibitem [{\citenamefont {Wehrle}(2022)}]{2729331}%
  \BibitemOpen
  \bibfield  {author} {\bibinfo {author} {\bibfnamefont {M.}~\bibnamefont
  {Wehrle}},\ }\href@noop {} {\emph {\bibinfo {title} {{Thermic FEA of TANB -
  LHCb Upgrade 2}}}},\ \bibinfo {type} {Tech. Rep.}\ \bibinfo {number} {2729331
  v.0.2}\ (\bibinfo  {institution} {CERN},\ \bibinfo {year} {2022})\BibitemShut
  {NoStop}%
\end{thebibliography}%

\end{document}